\documentclass[sigconf]{acmart}

\AtBeginDocument{%
  }

\copyrightyear{2026}
\acmYear{2026}
\setcopyright{cc}
\setcctype{by}
\acmConference[ICPE '26]{Proceedings of the 17th ACM/SPEC International Conference on Performance Engineering}{May 04--08, 2026}{Florence, Italy}
\acmBooktitle{Proceedings of the 17th ACM/SPEC International Conference on Performance Engineering (ICPE '26), May 04--08, 2026, Florence, Italy}
\acmDOI{10.1145/3777884.3797010}
\acmISBN{979-8-4007-2325-4/2026/05}

\usepackage{graphicx}
\usepackage{listings,pmboxdraw}
\usepackage{algorithm}
\usepackage{amsmath}
\usepackage{bbding}
\usepackage{amsthm}
\usepackage{multirow}
\usepackage{enumitem}

\usepackage[utf8]{inputenc}
\usepackage{xspace}

\newcommand\code[1]{{\tt \small #1}}

\lstnewenvironment{inlinelisting}{
\lstset{language=Java,
numbers=none,
basicstyle=\fontsize{8}{8}\selectfont\ttfamily
}}{}

\colorlet{punct}{red!60!black}
\definecolor{background}{HTML}{EEEEEE}
\definecolor{delim}{RGB}{20,105,176}
\colorlet{numb}{magenta!60!black}

\begin{document}

\title[MapReplay: Trace-Driven Benchmark Generation for Java HashMap]{MapReplay: Trace-Driven Benchmark Generation\\ for Java HashMap}

\author{Filippo Schiavio}
\email{filippo.schiavio@usi.ch}
\affiliation{
\institution{Università della Svizzera italiana}
\city{Lugano}
\country{Switzerland}}

\author{Andrea Rosà}
\email{andrea.rosa@usi.ch}
\affiliation{
\institution{Università della Svizzera italiana}
\city{Lugano}
\country{Switzerland}}

\author{Júnior Löff}
\email{loeffj@usi.ch}
\affiliation{
\institution{Università della Svizzera italiana}
\city{Lugano}
\country{Switzerland}}

\author{Lubomír Bulej}
\email{bulej@d3s.mff.cuni.cz}
\orcid{}
\affiliation{%
  \institution{Charles University}
  \city{Praha}
  \country{Czech Republic}
}

\author{Petr Tůma}
\email{petr.tuma@d3s.mff.cuni.cz}
\orcid{}
\affiliation{%
  \institution{Charles University}
  \city{Praha}
  \country{Czech Republic}
}
\author{Walter Binder}
\email{walter.binder@usi.ch}
\affiliation{
\institution{Università della Svizzera italiana}
\city{Lugano}
\country{Switzerland}}

\begin{abstract}
Hash-based maps, particularly \code{java.util.HashMap}, are pervasive in Java applications and the JVM, making their performance critical.
Evaluating optimizations is challenging because performance depends on factors such as operation patterns, key distributions, and resizing behavior. 
Microbenchmarks are fast and repeatable but often oversimplify workloads, failing to capture the realistic usage patterns. Application benchmarks (e.g., DaCapo, Renaissance) provide realistic usages but are more expensive to run, prone to variability, and dominated by non-\code{HashMap} computations, making map-related performance changes difficult to observe.

To address this challenge, we propose {\sf MapReplay}, a benchmarking methodology that combines the realism of application benchmarks with the efficiency of microbenchmarks.
{\sf MapReplay} traces \code{HashMap} API usages generating a \emph{replay workload} that reproduces the same operation sequence  while faithfully reconstructing internal map states.
This enables realistic and efficient evaluation of alternative implementations under realistic usage patterns. 
Applying {\sf MapReplay} to DaCapo-Chopin and Renaissance, the resulting suite---{\sf MapReplayBench}---reproduces application-level performance trends while reducing experimentation time and revealing insights difficult to obtain from full benchmarks.
\end{abstract}

\ccsdesc[500]{Software and its engineering~Software performance}

\keywords{Benchmarking, Tracing, Record/Replay, Hash-based Maps, Java Virtual Machine, Java Microbenchmark Harness}

\maketitle

\section{Introduction}\label{sec:introduction}

Hash-based maps are among the most widely used data structures in modern software systems~\cite{CollectionsBench:ICPE:2017}. Improving their performance and memory efficiency has been an active topic for decades, ranging from research on cache- and SIMD-efficient designs \cite{lim2014efficient,boether2023vectorized,bender2023iceberg}, storage- and persistence-aware hash tables \cite{10.5555/3291168.3291202,vogel2022plush}, to  industry-driven optimizations in widely deployed libraries and runtimes \cite{openjdkJEP180,abseilSwissTables,metaF14}. Moreover, hash-based maps are fundamental building blocks across a wide range of computing systems and techniques—e.g., database engines and data-management pipelines~\cite{boncz2020fsst}, data-driven compression and encoding schemes~\cite{jiang2021good}, and performance-critical runtime and storage systems that rely on carefully engineered hash tables~\cite{bender2023iceberg}. In Java, \code{java.util.HashMap}\footnote{In the following text, we omit fully qualified names for brevity.} is ubiquitous—employed in application code, standard libraries, and even within the Java Virtual Machine (JVM) itself. 
The performance of hash maps depends on numerous interrelated factors, including the mix and order of operations, the distribution of key hash codes, the current map size, and resizing behavior. These dependencies make it difficult to evaluate proposed optimizations reliably.

Microbenchmarks are commonly used to assess the performance of specific \code{HashMap} operations under controlled conditions. While they allow repeatable and fast experimentation, their workloads are often simplistic and fail to capture the complex access patterns that occur in real applications. Typically, distinct microbenchmarks are written for different operations---such as insertion, lookup, or iteration---each emphasizing a particular performance aspect~\cite{CollectionsBench:ICPE:2017}. Yet, when assessing a potential optimization, such as changing the default initial capacity or resizing strategy, it is unclear how to weigh improvements in one operation type against regressions in another. Consequently, microbenchmarks alone rarely provide a robust basis for deciding whether an optimization should be adopted in the standard library.

At the other end of the spectrum, full application benchmarks such as DaCapo Chopin~\cite{Blackburn2006,DaCapo-Chopin:ASPLOS:2025} and Renaissance~\cite{Prokopec2019} expose realistic \code{HashMap} behavior. However, they exhibit substantial measurement variability and are time-consuming to execute. Because the time spent in \code{HashMap} code represents only a small fraction of total execution time, observing statistically significant performance changes may require many repetitions---sometimes amounting to hundreds of hours of benchmarking---before meaningful conclusions can be drawn. Moreover, much of this effort is spent on non-map-related computation that are not key when evaluating \code{HashMap} alternatives.

In this paper, we propose a middle-ground approach that combines the representativeness of application benchmarks with the efficiency and control of microbenchmarks. We introduce \textbf{{\sf MapReplay}}, a tool that traces an application’s use of \code{HashMap} and offers a \textit{replay workload} reproducing the same sequence of map operations, but without any of the surrounding application logic. The replay workload faithfully reconstructs internal map states that are equivalent with respect to \code{HashMap}’s control-flow decisions, aiming to exercise the same \code{HashMap} code paths that are exercised as in the original application. This approach enables rapid, realistic, and reproducible evaluation of alternative map implementations under real-world workloads, without the cost and complexity of re-executing entire applications. Moreover, {\sf MapReplay} can produce a realistic, map-intensive replay workload from an application where map usage originally represents only a small portion of total execution time, thereby focusing the measurement effort on the performance aspects that matter most. Achieving faithful trace replay requires balancing fidelity and efficiency. {\sf MapReplay}’s design ensures accurate reproduction of performance-critical library code paths while minimizing trace size and replay overhead.

Thanks to {\sf MapReplay}, we obtain {\sf MapReplayBench}, a benchmark suite for hash-based Java maps created by applying our approach to two widely used Java application benchmark suites: DaCapo-Chopin~\cite{Blackburn2006,DaCapo-Chopin:ASPLOS:2025} and Renaissance~\cite{Prokopec2019}.
{\sf MapReplayBench} enables researchers and practitioners to analyze the performance of \code{HashMap}—as well as alternative hash-based map implementations on the JVM—under representative, application-derived workloads. We demonstrate the feasibility of {\sf MapReplay} by applying it to evaluate \code{HashMap} variants that differ in their default initial capacity. We use three benchmark classes: (1)~publicly available \code{HashMap} microbenchmarks, (2)~full application benchmarks from DaCapo Chopin and Renaissance, and (3)~our replay workloads. We show that {\sf MapReplay} provides performance insights consistent with map-intensive application benchmarks but at a smaller execution cost. Moreover, because {\sf MapReplay} amplifies the map-related activity of an application, it enables users to observe performance trends even in replay workloads derived from applications that are not map-intensive, and hence from which no clear conclusions could be drawn using the original application executions.

This paper makes the following contributions: 
\begin{enumerate}[noitemsep, leftmargin=*]\vspace{-.3mm}
  \item We present {\sf MapReplay}, a novel trace-driven benchmarking method for Java \code{HashMap}. 
  {\sf MapReplay} records the use of \code{HashMap} operations in applications and provides a replay workload that executes the same operation sequences under equivalent internal map states for the exercised \code{HashMap} code paths. 
  Its design reconciles \textit{fidelity}---faithfully reproducing performance-critical \code{HashMap} execution paths---with \textit{efficiency}, through compact tracing and lightweight replay.

  \item We use {\sf MapReplay} to generate traces from the established Java benchmark suites DaCapo-Chopin~\cite{Blackburn2006, DaCapo-Chopin:ASPLOS:2025} and Renaissance~\cite{Prokopec2019}.
  For these traces, our replay workloads show performance trends consistent with the application benchmarks while reducing experimentation time.
  Moreover, we demonstrate that replay workloads can reveal performance insights that may be difficult or impossible to obtain with the application benchmarks.
  
  \item We create {\sf MapReplayBench}, a stand-alone suite of ready-to-use replay workloads, consisting of traces generated from DaCapo-Chopin and Renaissance, and a replay infrastructure that executes them.\footnote{{\sf MapReplay} and {\sf MapReplayBench} are publicly available at \url{https://github.com/usi-dag/MapReplay}.} Thanks to {\sf MapReplayBench}, researchers and practitioners can evaluate alternative implementations of \code{HashMap} on map-intensive and realistic workloads.
\end{enumerate}

After providing background on \code{HashMap} (Sec.~\ref{sec:background}), we present the design and architecture of {\sf MapReplay} (Sec.~\ref{sec:design}), {\sf MapReplayBench} (Sec.~\ref{sec:MapReplayBench}), and our evaluation (Sec.~\ref{sec:eval}). 
Then, we discuss the broader implications, strengths, and limitations of our approach (Sec.~\ref{sec:discussion}), related work (Sec.~\ref{sec:relwork}), and conclude (Sec.~\ref{sec:conclusion}).

\section{Background: HashMap}\label{sec:background}

The \code{HashMap} class is the standard hash-based map implementation in the Java Class Library~\cite{JavaHashMapAPI25}.
Internally, it maintains a table of buckets, each containing zero or more keys, i.e., value pairs whose hash codes resolve to the same bucket index. The table is allocated lazily, meaning no memory is reserved until the first insertion occurs. When a key-value pair is added, the key’s hash code is combined with the table’s capacity (a power of two) to compute the bucket index using a bitwise mask, effectively equivalent to taking the hash code modulo the table capacity.

Each bucket is represented as a linked list of entries under normal conditions. As elements accumulate and the total number of mappings exceeds a threshold determined by the current capacity and load factor, the table is resized to double its previous capacity. The threshold ensures that the average bucket length remains small, maintaining expected constant-time performance for lookup, insertion, and removal operations. Notably, the table never shrinks, even after a call to \code{clear()}, to avoid costly reallocations in future insertions. The \textit{initial capacity} and \textit{load factor} parameters, which default to 16 and 0.75 respectively, control the initial table size and the resize threshold (i.e., $16 \cdot 0.75$).

In rare cases where many keys collide into the same bucket, the corresponding linked list may be transformed into a balanced red–black tree, a process known as \textit{treeification}~\cite{openjdkJEP180}, to mitigate performance degradation of lookup, insertion and removal from linear to logarithmic time. Conversely, if the number of entries in a treeified bucket decreases below a threshold, it is converted back into a linked list. These mechanisms collectively ensure that \code{HashMap} combines average-case constant-time operations with robust worst-case guarantees. In practice, treeification occurs only very rarely and is indicative of an unsuitable hash function.

\section{Design and Architecture}\label{sec:design}

This section presents the design and implementation principles of {\sf MapReplay}, our tool for trace-based benchmark synthesis targeting \code{HashMap}. We first present the terminology used in this paper and the overall software architecture of {\sf MapReplay}, then we motivate our high-level goals, describe our key design decisions, discuss performance considerations and outline the design and implementation of the tracer.

\subsection{Terminology and MapReplay Architecture}\label{sec:terminology}

\begin{figure}[t]
    \centering
    \includegraphics[width=\linewidth]{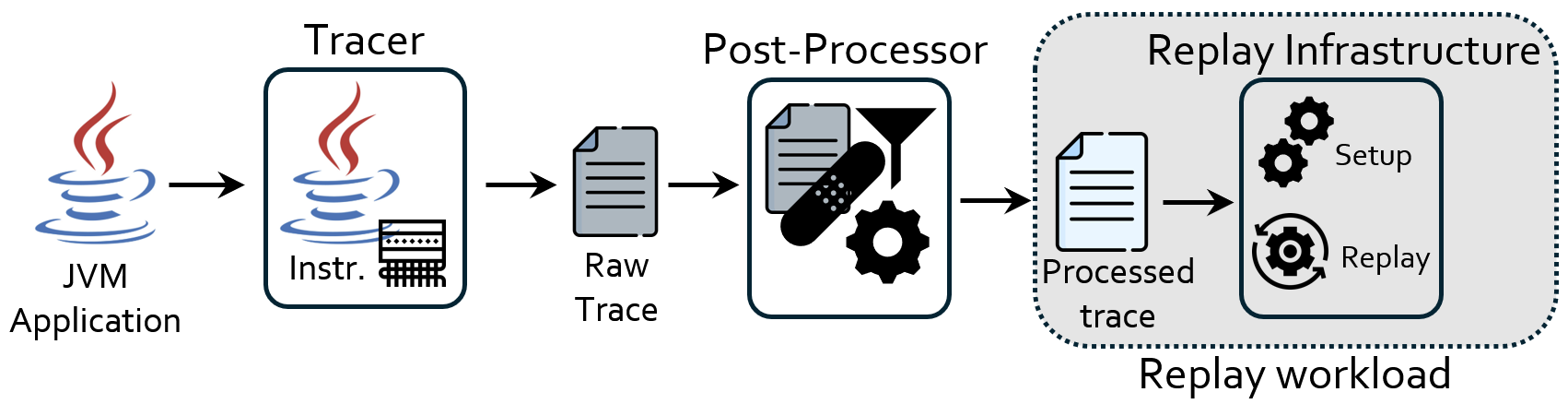}\vspace{-0.3cm}
    \caption{MapReplay architecture.}
    \label{fig:architecture}
\end{figure}

The {\sf MapReplay} system consists of three main components: the tracer, the offline trace post-processor, and the replay infrastructure. Figure~\ref{fig:architecture} illustrates their interactions.

\paragraph{\textbf{Tracer.}}
The \emph{tracer} instruments the application to capture relevant \code{HashMap} operations and their parameters. It records minimal information---such as operation type, target map, key identity, and hash code---to ensure low overhead during application execution. The output of the tracer is a \emph{raw trace}, i.e., a sequence of \code{HashMap}-relevant events. 

\paragraph{\textbf{Offline trace post-processor.}}
After tracing, the \emph{offline trace post-processor} sanitizes and compacts the raw trace. It removes incomplete map traces (e.g., without creation events), coalesces iterator operations, inserts object-free events, and encodes operations into a compact format suitable for efficient replay. The output of this step is a \emph{processed trace}. 

\paragraph{\textbf{Replay infrastructure and replay workload.}}
The final component is the \emph{replay infrastructure}, a standalone harness that interprets a processed trace and performs the corresponding \code{HashMap} operations. A \emph{replay workload} is a specific benchmark instance obtained by pairing the replay infrastructure with a processed trace. A replay workload consists of two phases: a \emph{setup} phase that initializes mockup objects and allocates data structures, and a \emph{replay} phase that executes the recorded sequence of map operations. The replay phase constitutes the benchmark method executed by performance-measurement frameworks (i.e., JMH~\cite{JMH} in the case of {\sf MapReplay}). Its single-threaded design and minimal runtime overhead ensure faithful performance evaluation of \code{HashMap} implementations. 

\subsection{Goals}

The central idea behind {\sf MapReplay} is to model an application as a sequence of \emph{map operations} interleaved with other, non-map computations. Our approach traces only the map-related operations, such as insertions, lookups, removals, and iterations, while disregarding all other application logic. The replay workload then re-executes the same sequence of map operations in equivalent map states, thereby reconstructing the internal control flow of the \code{HashMap} implementation as it occurred in the original application. This simplified model isolates the behavior of the map itself, enabling precise performance measurements focused solely on the library code. Naturally, this abstraction omits several factors that can influence performance in real executions, including concurrency and contention between threads, garbage-collection activity, dynamic compilation effects, and the trace-interpretation overhead of replay. Nonetheless, it provides a practical and effective foundation for realistic yet efficient evaluation of \code{HashMap} performance.

\paragraph{\textbf{Faithful execution paths.}}
The foremost goal of {\sf MapReplay} is to ensure that map operations in the replay workload---particularly those that are performance-critical---execute the same code paths within the library implementation as in the traced application. This is crucial to accurately evaluate proposed optimizations in the \code{HashMap} implementation under realistic conditions. Any deviation in control flow through library code could yield misleading performance assessments.

\paragraph{\textbf{Equivalent map state.}}
To reproduce the relevant \code{HashMap} execution paths faithfully, the internal state of each map at replay time must be equivalent to its state during the traced execution. Since many \code{HashMap} operations depend on structural properties such as bucket occupancy, collision chains, and resizing state, ensuring equivalence of the replayed map state is essential. Our tracing and replay mechanism therefore captures just enough information to reproduce the same evolution of the map state.

\subsection{Design Decisions}

Building on the design goals outlined above, {\sf MapReplay} incorporates several key decisions to balance fidelity, efficiency, and practical usability in trace collection and replay.

\paragraph{\textbf{Operation categorization.}}
We categorize map operations along two dimensions: (a) whether they are \textit{performance-critical} (requiring faithful replay of library code paths), and (b) whether they \textit{alter the map state} (thus influencing subsequent operations). This categorization defines the tracing and replay requirements:
\begin{itemize}[nosep, leftmargin=*]
  \item (a): Operations that are performance-critical (and may or may not be state-altering) are traced and faithfully replayed. Examples: \code{get()}, \code{put()}, \code{remove()}.
  \item ((not a) and b): Operations that are not performance-critical but may alter state are traced but may be replaced with equivalent state transitions during replay. Example: \code{compute()}.
  \item ((not a) and (not b)): Operations that are neither performance-critical nor state-altering are ignored. Example: \code{containsValue()}.
\end{itemize}
This two-dimensional classification provides the conceptual foundation for balancing fidelity and trace compactness.

\paragraph{\textbf{Substitution of application keys, minimized replay of callbacks.}}
Library code frequently invokes callbacks into application code via methods such as \code{hashCode()} and \code{equals()} on key objects (these primitives are themselves a target for SIMD-style acceleration on modern JVMs~\cite{loeff2024vectorizedIntrinsics,10.1145/3578360.3580265}). Additionally, some methods of the \code{HashMap} API (e.g., \code{compute()}) take user-defined functions or lambdas. Since {\sf MapReplay} focuses on benchmarking the library code rather than application logic, we abstract away user-defined behavior. We record only essential properties of keys (in particular, their hash codes) and substitute all keys with a uniform mockup-key type in the replay environment. This design drastically reduces trace size and avoids complex dependencies on application classes.

\paragraph{\textbf{Preserving hash collisions.}}
To maintain equivalence in map structure, we accurately preserve the hash codes of all keys. This ensures that during replay, the same hash-bucket assignments and collision chains occur, leading to the same library-level control flow.

\paragraph{\textbf{Simplified equality semantics.}}
Preserving key equality relations would require expensive tracking of equivalence classes. Instead, {\sf MapReplay} enforces identity-based equality: when an operation references a key equal (but not identical) to an existing one, the trace records the existing key instead. This guarantees that replay can use trivial \code{equals()} implementations for mockup keys while still reproducing the correct behavior and control flow in the map.

\paragraph{\textbf{Ignoring values.}}
We do not preserve any information about values stored in the map. Operations such as \code{containsValue()} are rare (and inefficient) in practice, and the state of objects stored as values in maps may be modified directly by the application, which would require tracing arbitrary field writes.\footnote{Note that this problem does not arise for keys in a map. The Java documentation states that "the behavior of a map is not specified if the value of an object is changed in a manner that affects equals comparisons while the object is a key in the map"~\cite{JavaHashMapAPI25}. Hence, we assume that the equality relation remains stable for all keys.} Therefore, such operations are excluded from tracing and replay.

\paragraph{\textbf{Tracing iterators and spliterators.}}
Operations involving map views---specifically iterators and spliterators---are traced because they are widely used in real applications and often performance-relevant. These operations are included in replay to accurately reflect iteration and traversal patterns.

\paragraph{\textbf{Exception handling.}}
Operations that throw exceptions in applications (e.g., \code{ConcurrentModificationException}) are typically rare and do not modify map state (they are indicative of application bugs). Consequently, we omit tracing and replay of exceptional operations, allowing the replay infrastructure to assume exception-free execution. This simplifies the replay workload and avoids unnecessary control-flow overhead.

\paragraph{\textbf{No Hash-Map polymorphism.}}
To ensure consistent and comparable results, {\sf MapReplay} enforces that all hash maps instantiated in a benchmark run are of the same type. This design choice guarantees that every map operation in the replay workload is invoked from a monomorphic call site, allowing the JIT compiler to apply aggressive optimizations such as inlining. The resulting compiled code reduces the interpretation overhead of replay and ensures that most of the benchmark time is spent executing the library code under evaluation. For simplicity, {\sf MapReplay} does not currently distinguish between \code{HashMap} subtypes during tracing; for instance, \code{LinkedHashMap} instances in the original application are not replayed separately. Extending {\sf MapReplay} to also trace specific subtypes such as \code{LinkedHashMap} would, however, be straightforward.

\paragraph{\textbf{Offline trace sanitization.}}
To minimize the runtime overhead of tracing, instrumentation remains simple. Expensive cleanup and consistency checks occur in an offline post-processing phase. This includes removing map operations for which creation events could not be captured, such as maps created during JVM bootstrapping, when tracing is not yet possible.\footnote{While \code{HashMap} instances are created already in an early phase of JVM bootstrapping, even a tracer written in native code (as in our implementation) could not be invoked, because the use of JNI would crash bootstrapping~\cite{10.1145/2414740.2414747}.} Excluding these operations ensures replay consistency.

\paragraph{\textbf{Single-threaded replay.}}
The replay workload is single-threaded, even if the original application used multiple threads. If threads operate on distinct \code{HashMap} instances, their operations may be safely reordered. For immutable shared maps, read-only operations may also be reordered. For mutable shared maps, we assume proper synchronization in the original application, preserving the relative order of map operations within critical sections. This reasoning is valid for non-thread-safe maps like \code{HashMap} but not for concurrent collections such as \code{ConcurrentHashMap}, which we discuss in Sec.~\ref{sec:limitations}.

\subsection{Performance Considerations}

To ensure that the replay workloads accurately reflect library-level performance, {\sf MapReplay}’s design ensures compact traces as well as minimized replay overhead, while preserving faithful execution of \code{HashMap} operations.

\paragraph{\textbf{Efficient replay.}}
The replay mechanism must impose negligible overhead relative to the execution of library code. Therefore, the benchmark method itself is designed to perform minimal decoding and dispatch work, keeping the focus on the performance of the map operations under test.

\paragraph{\textbf{Coalescing iterator operations.}}
In many traces, iterator methods such as \code{hasNext()} and \code{next()} occur extremely frequently and are individually inexpensive. Replaying them one-by-one would make replay overhead dominant. During trace post-processing, we therefore coalesce consecutive iterator or spliterator operations (e.g., \code{tryAdvance()}) into aggregated replay events, unless interrupted by mutating operations such as \code{remove()}. The benchmark replays these as short loops executing the equivalent number of steps.

\paragraph{\textbf{Compact runtime encoding.}}
To reduce interpretation overhead, we represent the replay trace using primitive arrays that encode operations as opcodes with operands (e.g., map IDs). Arrays store active \code{HashMap} instances, iterators, and spliterators. This low-level encoding provides fast decoding and dispatch during replay.

\paragraph{\textbf{Minimized object allocation.}}
All benchmark-related objects---including mockup keys and data structures---are preallocated during the benchmark setup phase. During replay, only necessary runtime objects such as maps and iterators are created. This reduces unrelated allocation and garbage collection activity during replay, which can reduce measurement noise when isolating \code{HashMap}-local effects.

\paragraph{\textbf{Explicit object-free events.}}
To further reduce memory footprint and GC overhead, the trace includes explicit object-free events emitted immediately after the last use of an object. These events nullify references to allow early reclamation by the garbage collector. Since these events are inserted in the offline post-processing phase, they hardly affect replay overhead but can help keep automated memory management efficient during repeated trace replay.

\subsection{Tracer Design and Implementation}\label{sec:tracer}

Here, we discuss the main design choices regarding {\sf MapReplay}'s tracer and describe its implementation details. 
 
\paragraph{\textbf{Source-level instrumentation and deployment.}}
To collect traces with minimal disturbance, {\sf MapReplay} instruments \code{HashMap} by directly modifying its Java source code at carefully chosen internal points. This instrumentation records \code{HashMap}-relevant events (e.g., public operations and selected internal outcomes) while preserving the original control flow and data-structure semantics. Recompiling the JVM to use the instrumented \code{HashMap} is not required: as suggested by related work~\cite{misleading-micro}, we deploy the instrumented \code{HashMap} by patching the \code{java.base} module at launch time (via the module-patching mechanism~\cite{module-patching} available since Java~9), such that the application executes our instrumented class in place of the standard library version.

\paragraph{\textbf{JNI-based trace recording.}}
The inserted instrumentation probes are intentionally small and immediately delegate to trace-recording routines implemented as native methods. Concretely, each probe performs only lightweight argument preparation (e.g., computing key identifiers) and then invokes (via JNI~\cite{jni}) a native method that appends the corresponding event to a raw trace buffer. Implementing the recording logic in native code keeps the Java-side probes simple and avoids allocations and library calls, while still enabling efficient I/O and compact binary encoding of events.

\paragraph{\textbf{Avoiding instrumentation reentrancy.}}
A key challenge is preventing \emph{instrumentation reentrancy}: if the tracing or profiling logic itself executes instrumented library code, the probes can re-trigger during the execution of the instrumentation, leading to unbounded recursion and eventual failure (i.e., \code{StackOverflowError}). This issue is particularly acute when instrumenting core library classes such as \code{HashMap}, because they are used pervasively throughout the Java Class Library; as a result, seemingly unrelated functionality needed for trace collection (e.g., initialization of common Java classes that support I/O writing) may indirectly allocate or access \code{HashMap} instances. Prior work on JVM observability highlights that this kind of self-interference is a practical obstacle to reliable library instrumentation and often requires careful separation between application execution and analysis logic~\cite{10.1145/2414740.2414747}. To robustly avoid reentrancy in {\sf MapReplay}, we keep the Java-side probes minimal and implement the trace-recording logic in native code, relying on C libraries rather than the Java Class Library.

\paragraph{\textbf{Alternatives to source-level instrumentation.}}
An alternative to source-level instrumentation is to instrument \code{HashMap} using load-time bytecode transformation or instrumentation frameworks inspired to Aspect-Oriented Programming principles (AOP). In practice, instrumenting the Java Class Library is often fragile: many bytecode instrumentation frameworks primarily target application classes rather than library ones, and instrumenting classes loaded in the early stage of the JVM lifecycle (such as \code{HashMap})  can interfere with JVM initialization or trigger self-interference problems unless special care is taken~\cite{10.1145/2414740.2414747}. There are, however, frameworks that explicitly support robust bytecode instrumentation of core classes. For example, DiSL~\cite{10.1145/2162049.2162077,rosa2017rsi,rosa2018rsi} provides full bytecode coverage and includes a \emph{dynamic bypass} mechanism based on polymorphic bytecode instrumentation~\cite{10.1145/1960275.1960292, 10.1002/spe.2385} that prevents instrumentation reentrancy by ensuring that analysis code executes against an uninstrumented version of selected classes while the application executes the instrumented version.

We did not adopt such a framework for {\sf MapReplay} for three reasons. First, our probes must be placed at multiple fine-grained internal locations inside \code{HashMap} (including mid-method points that are control-flow-specific) and must sometimes behave differently depending on the map state; encoding these semantics with an instrumentation framework---which often implies expressing the semantics with a limited domain-specific language---would be substantially more complex and error-prone than direct source modifications. Second, several public \code{HashMap} entry points delegate to each other (e.g., convenience overloads and constructor chaining), and we must avoid recording the same logical operation multiple times; performing the appropriate instrumentation is straightforward with manual source edits but difficult to enforce reliably with generic join-point-based instrumentation required by an AOP framework. Third, framework-based approaches introduce additional runtime complexity and overhead---e.g., class transformation during startup and, for some designs such as in DiSL, auxiliary processes and an increased memory footprint~\cite{rosa2018tgp,rosa2019tgp}.
---which is undesirable when collecting traces intended to faithfully reflect baseline library behavior. Overall, we found that direct source-level instrumentation combined with JNI-based recording provides the simplest and most reliable way to collect \code{HashMap} traces while keeping both measurement perturbation and resource consumption low.

\section{MapReplayBench}\label{sec:MapReplayBench}

In this section we present {\sf MapReplayBench}, a benchmark suite of replay workloads for hash-based Java maps created by applying {\sf MapReplay} to two widely used Java application benchmark suites: DaCapo-Chopin~\cite{Blackburn2006, DaCapo-Chopin:ASPLOS:2025} and Renaissance~\cite{Prokopec2019}.
{\sf MapReplayBench} enables researchers and practitioners to analyze the performance of the \code{HashMap} class---as well as alternative hash-based map implementations on the JVM---under representative, application-derived workloads.

\begin{table*}[t]
    \caption{HashMap usage in DaCapo-Chopin (C-) and Renaissance (R-) benchmarks. Percentage CPU time spent in HashMap stems from executing the original benchmarks. The other metrics stem from the MapReplay processed traces.}
    \label{tab:characterization}
    \centering
    {\footnotesize
\setlength{\tabcolsep}{2.5pt} 
\begin{tabular}{lrrrrrrr | lrrrrrrr} 
& \% CPU & & & & & & & & \% CPU &\\ 
Benchmark & in HM & \# Event & \# Create & \# Read  & \# Write & \# Iterate & Size & Benchmark & in HM & \# Event & \# Create & \# Read & \# Write & \# Iterate  & Size \\ \hline
R-par-mnemonics & 27.43 & 66236024 & 1415980 & 43872832 & 18678894 & 1252797 & 102M & R-naive-bayes & 0.28 & 501175 & 57466 & 264564 & 140864 & 15823 & 1.1M \\
R-mnemonics & 26.00 & 66203404 & 1415852 & 43872832 & 18646430 & 1252783 & 60M & C-batik & 0.27 & 396511 & 2031 & 338706 & 52196 & 1769 & 400K \\
C-biojava & 23.49 & 394224921 & 150337 & 364892378 & 29031647 & 43352 & 258M & C-jython & 0.21 & 2563693 & 540457 & 1939448 & 72575 & 9763 & 2.0M \\
R-scrabble & 16.44 & 9600913 & 1244173 & 672755 & 5197588 & 1243212 & 33M & R-gauss-mix & 0.11 & 497218 & 75100 & 214081 & 129394 & 11954 & 1.4M \\
R-scala-stm-bench7 & 13.67 & 49156302 & 35940 & 42071490 & 7013480 & 185 & 222M & R-dec-tree & 0.10 & 495801 & 47258 & 311509 & 104307 & 9659 & 1.1M \\
R-fj-kmeans & 6.76 & 102274181 & 411953 & 6565 & 101033413 & 410939 & 200M & R-als & 0.10 & 503414 & 65069 & 255852 & 131942 & 7855 & 1.4M \\
C-h2 & 5.38 & 139423811 & 7242709 & 67714731 & 45978767 & 12938167 & 435M & C-luindex & 0.08 & 2393138 & 1001 & 2388438 & 2655 & 559 & 232K \\
R-rx-scrabble & 3.87 & 1746158 & 50357 & 63296 & 1533327 & 49593 & 6.8M & R-neo4j-analytics & 0.08 & 8078074 & 593295 & 2930709 & 3411231 & 570423 & 3.8M \\
C-tomcat & 3.34 & 13220831 & 70100 & 6762075 & 5924407 & 317981 & 59M & R-chi-square & 0.02 & 111593 & 12953 & 53919 & 37082 & 2809 & 256K \\
C-kafka & 3.16 & 12116302 & 67034 & 8166290 & 3503230 & 328965 & 41M & C-zxing & 0.02 & 578096 & 23171 & 235569 & 254232 & 42651 & 1.1M \\
C-pmd & 2.93 & 15514296 & 1921133 & 6116013 & 4252211 & 2908055 & 68M & R-page-rank & 0.02 & 127214 & 16085 & 57973 & 41032 & 5498 & 304K \\
C-xalan & 0.98 & 3086931 & 74250 & 2419000 & 534399 & 17597 & 9.0M & R-dotty & 0.02 & 27815 & 9816 & 9054 & 8576 & 171 & 80K \\
C-spring & 0.91 & 7762498 & 431686 & 5040616 & 1524973 & 462283 & 28M & C-jme & 0.02 & 11189 & 911 & 4627 & 4946 & 366 & 36K \\
C-fop & 0.69 & 937499 & 40275 & 759269 & 94830 & 19579 & 976K & R-akka-uct & 0.01 & 157637 & 5932 & 80118 & 60358 & 6895 & 460K \\
C-avrora & 0.61 & 2513654 & 319 & 2509962 & 3019 & 161 & 1.6M & C-graphchi & 0.00 & 14595 & 2180 & 4016 & 5987 & 508 & 44K \\
C-eclipse & 0.61 & 4089948 & 116315 & 2762388 & 1030205 & 93647 & 8.5M & R-scala-doku & 0.00 & 16087 & 1087 & 6736 & 8013 & 139 & 52K \\
R-movie-lens & 0.49 & 2909774 & 401752 & 1303789 & 788540 & 36264 & 8.7M & R-reactors & 0.00 & 16651 & 1487 & 6898 & 7869 & 216 & 52K \\
R-log-regression & 0.46 & 619471 & 78414 & 320235 & 161589 & 9802 & 1.6M & C-sunflow & 0.00 & 3293 & 376 & 474 & 2138 & 167 & 16K \\
R-finagle-http & 0.42 & 3141156 & 18302 & 3102359 & 18844 & 539 & 8.5M & R-future-genetic & 0.00 & 10285 & 1117 & 3594 & 5323 & 139 & 32K \\
C-h2o & 0.37 & 615037 & 26745 & 440792 & 114519 & 9474 & 1.1M & R-philosophers & 0.00 & 15938 & 954 & 6713 & 8017 & 140 & 52K \\
C-lusearch & 0.36 & 7956080 & 48572 & 7732258 & 99144 & 34982 & 18M & R-scala-kmeans & 0.00 & 15554 & 722 & 6574 & 8009 & 138 & 52K \\
R-finagle-chirper & 0.28 & 2598361 & 5806 & 2576093 & 15241 & 357 & 6.4M \\

\hline
\end{tabular}}

\end{table*}

Not all benchmarks in the two suites exhibit substantial map activity.
Table~\ref{tab:characterization} characterizes the map usage of each benchmark.
The first column, “\% CPU in HM,” reports the percentage of CPU time originally spent inside the \code{HashMap} class; we use this metric to define the map intensity of a workload, and the table is sorted accordingly.
The remaining columns describe the events collected by the {\sf MapReplay} tracer and refined by the offline post-processor.

The “\# Event” column reports the total number of replay events after post-processing.
The “\# Create” column counts all invocations of \code{HashMap} constructors.
“\# Read” covers all traced public methods that inspect the state of the map (e.g., \code{get}, \code{containsKey}), while “\# Write” includes all methods that modify the map (e.g., \code{put}, \code{remove}).
“\# Iterate” summarizes operations that traverse the map or its views, including iterators and spliterators over entries, keys, or values.
Finally, the “Size” column reports the size in bytes of the processed trace.\footnote{Each trace is a compressed archive. Traces may have similar event numbers, but can significantly differ in size (e.g., mnemonics and par-mnemonics) due to compression.}

\section{Evaluation}\label{sec:eval}

Here, we evaluate {\sf MapReplay} against existing microbenchmark and application-benchmark methodologies.
We show their limitations focusing on a simple yet representative case study: How effective are current benchmarking approaches in helping developers determine the optimal \emph{default initial capacity (DIC)} of \code{HashMap}?

We design an experiment to assess whether increasing the DIC from the default value of 16 to 32, 64, or 128 yields measurable performance benefits.
Increasing the initial capacity can reduce resizing frequency, shorten collision chains, and accelerate unsuccessful lookups by increasing the likelihood of probing an empty bucket. 
However, larger tables incur higher allocation costs, slower iterations over sparsely populated buckets, and reduced locality.
The net effect of these trade-offs is highly application-dependent.
Our objective is to compare {\sf MapReplay} against microbenchmarks and application benchmarks in determining whether higher DIC values improve performance.

\subsection{Experimental Setup}
We conduct our evaluation on a machine with a 16-core Intel Xeon Gold 6326 CPU @~2.9GHz and 256GB of RAM @~3.2GHz, running Ubuntu 22.04 LTS (kernel 5.15.0-25-generic).
Hyper-threading and turbo boost are disabled to ensure stable and reproducible performance measurements.
All experiments use JDK~25 (Oracle JDK~25, build 25.0.1+8-LTS-27).

For each case study, we rely on the latest version of JMH available at the time of writing (version~1.37), carefully following best practices to avoid common benchmarking pitfalls~\cite{8747433}.
Each JMH configuration consists of 5 JVM runs, each with 5 warmup iterations followed by 5 measured iterations.
For the microbenchmarks, we set the iteration duration to 500ms, as individual benchmark invocations execute quickly (typically from a few ns to 1~ms).
For the replay workloads, we use an iteration duration of 10s, since a single invocation may take significantly longer (up to approximately 3.5s).
For application benchmarks, we execute 30--100 iterations (depending on the workload) to ensure adequate warmup, and collect results from 30 JVM runs to mitigate measurement noise.

\begin{figure*}[t]
\centering
\includegraphics[width=\textwidth]{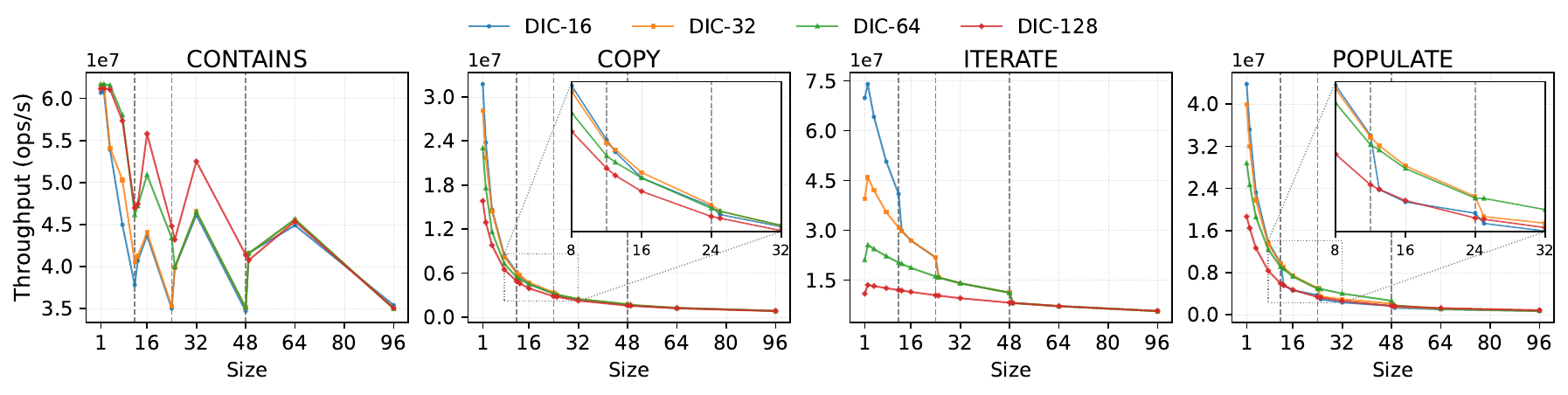}
\caption{Microbenchmark performance for different operations, map sizes, and DIC configurations. Vertical gray lines indicate thresholds where resizing occurs.}
\label{fig:micros}
\end{figure*}

\subsection{Microbenchmarks}\label{eval:micro}

Here we demonstrate the limitations of microbenchmarks, which often do not provide sufficient information to decide whether a given optimization should be incorporated in a standard library. To this end, we use an existing microbenchmark suite designed to study the performance of Java Collections~\cite{CollectionsBench:ICPE:2017}. We frame a performance evaluation around our case study to assess which DIC value offers the best performance for \code{HashMap}. Specifically, we measure performance across varying input sizes, with particular attention to sizes near thresholds that trigger resizing operations.

Focusing exclusively on execution time and ignoring secondary aspects such as memory consumption, we observe that performance depends strongly on the composition of operations, the number of operations performed at each map size, and the choice of key distribution. Notably, due to their synthetic nature, the microbenchmarks employ three fixed key distributions, perform a time-bounded amount of work per map size, and maintain a uniform composition of four equally weighted operations (i.e., \code{contains}, \code{copy}, \code{iterate}, and \code{populate}).

Fig.~\ref{fig:micros} reports the observed throughput on these four operations for different DIC settings.
The \code{contains} and \code{populate} operations generally benefit from larger DIC, as a larger table helps reduce collisions. For \code{contains}, lookup latency is directly tied to collision-chain length: maps with DIC-128 outperform smaller configurations until all maps reach comparable internal sizes (i.e., $\geq$ 49 elements). Performance often increases after each resize event (e.g., at 13, 25, and 49 elements). Some variability remains, due to random key selection in each execution~\cite{CollectionsBench:ICPE:2017}, though the benchmark ensures determinism across runs and DIC configurations.

For \code{populate}, each iteration allocates a new map. A larger DIC improves performance only when the number of inserted elements is sufficiently high; otherwise, mid-sized tables offer a better balance between allocation cost and collision rate.

In contrast, the \code{copy} and \code{iterate} operations tend to degrade with larger DIC. Both traverse all buckets, incurring additional overhead when the underlying table is sparse. The effect of resizing is particularly visible in \code{iterate}: each resize doubles the bucket array, causing the performance of smaller DIC configurations to align with the next larger configuration. Once all variants reach an internal capacity of 128 (at $\geq$ 49 elements), their performance converges. A similar trend appears in \code{copy}, although its allocation-heavy nature makes trends harder to interpret.

Based on the microbenchmarks, it is very hard to determine which DIC value to use in the library. The optimal choice depends on factors such as the distribution of map sizes, the relative frequency of operations, and the key distribution---factors that are synthetic in the microbenchmarks and may not reflect real application behavior. {\sf MapReplay} overcomes these limitations by tracing and replaying realistic map usage drawn from real workloads.

\subsection{Application Benchmarks Classification}\label{eval:classification}

To assess how different workloads react to DIC changes, we classify all benchmarks according to the percentage of execution time spent in \code{HashMap} operations\footnote{We collected CPU time in \code{HashMap} using profhot~\cite{profhot}.} (column ``\% CPU in HM'' in Table~\ref{tab:characterization}).
This metric reflects how intensively each workload exercises \code{HashMap} and therefore how sensitive it may be to DIC changes.
We distinguish three categories:

\paragraph{\textbf{Map-intensive (\boldmath$\%$ CPU in HM $\geq 5$.)}}
These benchmarks spend a substantial portion of their execution time in \code{HashMap} and are expected to be sensitive to DIC changes.

\paragraph{\textbf{Map-moderate (\boldmath$\%$ CPU in HM $< 5$ and \#Event $\geq 100{,}000$).}}
These workloads use \code{HashMap} non-trivially, but map operations are not the dominant cost.
Their performance is generally less affected by DIC changes, as \code{HashMap} operations represent only a small part of the overall execution.

\paragraph{\textbf{Map-minimal (\boldmath$\%$ CPU in HM $< 5$ and \#Event $< 100{,}000$).}}
These workloads exhibit negligible \code{HashMap} activity.
DIC modifications generally have no measurable impact, and the collected traces are often too small to be representative.
Consequently, we exclude these benchmarks from our evaluation. As shown in Table~\ref{tab:characterization}, their trace sizes typically amount to only a few kilobytes.

\subsection{Application Benchmarks}\label{eval:app}

Table~\ref{tab:eval:sizes} (left) reports execution times and speedup factors for different DIC configurations across the DaCapo-Chopin and Renaissance benchmark suites.
Most application benchmarks (excluding the few in the map-intensive category) show no measurable performance change when varying the DIC. This is expected---many workloads are not map-intensive (as shown in Table~\ref{tab:characterization}).

In contrast, nearly all benchmarks in the map-intensive category show performance differences when varying the DIC (with the exception of R-fj-kmeans). A few map-moderate benchmarks also show sensitivity---most notably R-rx-scrabble.
However, the small number of benchmarks showing meaningful differences makes it difficult to determine the ideal DIC configuration.

Furthermore, detecting small but statistically significant performance changes in the application benchmarks requires collecting measurements from many runs, each of which needs to execute long enough to get past the initial warm-up phase associated with the compilation of large amounts of code.
This requires developers to wait long before receiving feedback on configuration changes.
In this particular case, we collected data from 30 runs, with benchmark-specific repetition numbers, amounting to roughly 72 hours per DIC configuration.

\subsection{MapReplayBench}\label{eval:mapreplay}

\begin{table*}[t]
    \centering
    
    \caption{Performance comparison between application benchmarks and replay workloads (execution time in ms/op) for DaCapo-Chopin (C-) and Renaissance (R-) for different DIC values. Benchmarks above the horizontal line are \emph{map-intensive} ($\%$ CPU in HM $\geq 5$), while the ones below are \emph{map-moderate}. Benchmarks from the \emph{map-minimal} category were omitted as irrelevant.}
    \label{tab:eval:sizes}
    
    \renewcommand{\arraystretch}{1.1}
    \setlength{\tabcolsep}{4pt}

    {\footnotesize 
\begin{tabular}{l|rrrr|rrrr}
\hline
 & \multicolumn{4}{c|}{Application Benchmark} & \multicolumn{4}{c}{Replay Workloads} \\ \hline
Benchmark & DIC-16 & DIC-32 & DIC-64 & DIC-128 & DIC-16 & DIC-32 & DIC-64 & DIC-128 \\ \hline
 R-par-mnemonics  & 2051±25.7 & 2050±29.0 (1.00x) & 2007±26.9 (1.02x) & 2069±31.9 (0.99x) & 958±38.2 & 928±37.7 (1.03x) & 878±40.5 (1.09x) & 885±22.6 (1.08x) \\ 
 R-mnemonics  & 2499±28.6 & 2473±26.3 (1.01x) & 2420±19.3 (1.03x) & 2532±35.9 (0.99x) & 1022±101.8 & 979±56.4 (1.04x) & 932±10.0 (1.10x) & 963±53.2 (1.06x) \\ 
 C-biojava  & 8099±22.0 & 8062±24.2 (1.00x) & 7944±32.8 (1.02x) & 7924±31.1 (1.02x) & 3781±861.3 & 3592±137.4 (1.05x) & 3344±59.4 (1.13x) & 3349±137.7 (1.13x) \\ 
 R-scrabble  & 59±0.5 & 61±0.6 (0.97x) & 67±1.4 (0.88x) & 80±3.2 (0.73x) & 331±2.8 & 358±5.5 (0.92x) & 378±8.4 (0.88x) & 463±8.2 (0.72x) \\ 
 R-scala-stm-bench7  & 758±15.9 & 756±14.1 (1.00x) & 744±10.3 (1.02x) & 748±10.9 (1.01x) & 1006±8.3 & 982±41.5 (1.02x) & 986±4.6 (1.02x) & 953±32.8 (1.06x) \\ 
 R-fj-kmeans  & 988±2.3 & 990±3.5 (1.00x) & 985±3.0 (1.00x) & 987±3.0 (1.00x) & 1730±68.8 & 1600±142.1 (1.08x) & 1460±188.1 (1.18x) & 1559±57.4 (1.11x) \\ 
 C-h2  & 2437±8.9 & 2435±11.9 (1.00x) & 2452±12.7 (0.99x) & 2480±11.2 (0.98x) & 3065±139.6 & 3034±43.3 (1.01x) & 3086±316.6 (0.99x) & 3170±56.4 (0.97x) \\ 
\hline 
 R-rx-scrabble  & 102±0.7 & 102±0.7 (1.00x) & 103±0.7 (0.99x) & 107±0.6 (0.95x) & 67±1.0 & 67±1.5 (1.01x) & 69±0.8 (0.98x) & 69±1.9 (0.98x) \\ 
 C-tomcat  & 6809±1.7 & 6809±2.1 (1.00x) & 6810±1.6 (1.00x) & 6811±1.6 (1.00x) & 398±8.5 & 377±21.2 (1.05x) & 364±5.9 (1.09x) & 368±22.8 (1.08x) \\ 
 C-kafka  & 5014±0.9 & 5015±1.3 (1.00x) & 5015±1.2 (1.00x) & 5014±1.0 (1.00x) & 203±10.9 & 198±9.3 (1.02x) & 184±6.8 (1.10x) & 188±7.6 (1.08x) \\ 
 C-pmd  & 1395±7.9 & 1399±10.1 (1.00x) & 1393±8.9 (1.00x) & 1394±12.2 (1.00x) & 489±26.7 & 495±22.2 (0.99x) & 483±25.7 (1.01x) & 473±10.0 (1.04x) \\ 
 C-xalan  & 517±2.6 & 514±3.1 (1.00x) & 517±2.7 (1.00x) & 516±3.1 (1.00x) & 100±1.8 & 106±3.3 (0.94x) & 103±5.2 (0.97x) & 96±1.9 (1.04x) \\ 
 C-spring  & 1940±35.9 & 1941±38.3 (1.00x) & 1931±48.6 (1.00x) & 1938±40.1 (1.00x) & 201±5.2 & 178±5.0 (1.13x) & 182±12.2 (1.11x) & 184±1.7 (1.09x) \\ 
 C-fop  & 487±1.0 & 487±0.7 (1.00x) & 486±0.9 (1.00x) & 488±1.5 (1.00x) & 7±0.4 & 6±0.4 (1.15x) & 6±0.1 (1.11x) & 6±0.3 (1.12x) \\ 
 C-avrora  & 3673±6.0 & 3673±4.3 (1.00x) & 3673±3.7 (1.00x) & 3671±5.6 (1.00x) & 13±0.4 & 12±0.1 (1.10x) & 10±0.0 (1.24x) & 10±0.1 (1.23x) \\ 
 C-eclipse  & 12337±18.1 & 12333±21.6 (1.00x) & 12339±23.5 (1.00x) & 12336±21.2 (1.00x) & 54±10.5 & 61±7.5 (0.89x) & 62±13.3 (0.88x) & 61±10.2 (0.89x) \\ 
 R-movie-lens  & 6793±15.0 & 6806±14.5 (1.00x) & 6809±19.9 (1.00x) & 6809±13.8 (1.00x) & 56±0.3 & 52±2.3 (1.07x) & 51±3.5 (1.10x) & 51±3.6 (1.10x) \\ 
 R-log-regression  & 561±0.9 & 566±1.1 (0.99x) & 561±0.7 (1.00x) & 561±1.1 (1.00x) & 12±0.6 & 12±0.6 (1.02x) & 12±0.5 (1.03x) & 12±0.2 (1.04x) \\ 
 R-finagle-http  & 2009±8.9 & 2007±7.5 (1.00x) & 2002±8.0 (1.00x) & 2005±7.6 (1.00x) & 16±3.3 & 15±0.2 (1.05x) & 16±3.2 (0.98x) & 17±2.6 (0.94x) \\ 
 C-h2o  & 2942±11.9 & 2929±14.9 (1.00x) & 2937±20.2 (1.00x) & 2934±18.9 (1.00x) & 7±2.3 & 6±1.6 (1.07x) & 7±1.3 (1.03x) & 6±1.4 (1.16x) \\ 
 C-lusearch  & 1830±19.7 & 1838±12.2 (1.00x) & 1829±17.4 (1.00x) & 1829±17.7 (1.00x) & 63±1.8 & 63±1.0 (1.00x) & 58±20.6 (1.08x) & 65±2.7 (0.97x) \\ 
 R-finagle-chirper  & 1597±9.2 & 1598±5.9 (1.00x) & 1592±5.7 (1.00x) & 1594±7.2 (1.00x) & 14±0.3 & 14±0.3 (0.99x) & 14±0.2 (0.98x) & 14±0.1 (0.97x) \\ 
 R-naive-bayes  & 256±1.9 & 255±2.2 (1.00x) & 257±3.1 (0.99x) & 257±6.4 (1.00x) & 10±0.8 & 10±1.1 (1.01x) & 10±1.0 (1.01x) & 10±0.7 (1.02x) \\ 
 C-batik  & 1670±1.4 & 1671±2.2 (1.00x) & 1670±0.9 (1.00x) & 1670±1.3 (1.00x) & 4±0.2 & 4±0.1 (1.01x) & 4±0.1 (1.05x) & 4±0.3 (1.03x) \\ 
 C-jython  & 3894±83.0 & 3866±6.5 (1.01x) & 3868±6.4 (1.01x) & 3866±7.5 (1.01x) & 18±0.3 & 18±0.3 (1.00x) & 18±0.8 (0.99x) & 18±0.9 (0.98x) \\ 
 R-gauss-mix  & 4928±18.3 & 4933±13.5 (1.00x) & 4924±16.1 (1.00x) & 4922±18.3 (1.00x) & 11±0.2 & 11±0.4 (1.00x) & 11±0.2 (0.98x) & 11±0.4 (0.96x) \\ 
 R-dec-tree  & 631±4.7 & 632±4.8 (1.00x) & 631±4.8 (1.00x) & 632±4.6 (1.00x) & 9±0.5 & 9±0.5 (1.00x) & 9±0.1 (1.02x) & 9±0.8 (1.03x) \\ 
 R-als  & 1337±1.5 & 1337±2.5 (1.00x) & 1336±1.2 (1.00x) & 1338±2.0 (1.00x) & 11±0.5 & 10±0.7 (1.06x) & 10±0.1 (1.07x) & 10±0.6 (1.10x) \\ 
 C-luindex  & 4643±20.7 & 4654±21.6 (1.00x) & 4646±21.7 (1.00x) & 4643±15.2 (1.00x) & 10±0.0 & 10±0.2 (1.02x) & 10±0.3 (0.99x) & 10±0.0 (1.00x) \\ 
 R-neo4j-analytics  & 1229±8.2 & 1227±7.3 (1.00x) & 1226±8.3 (1.00x) & 1226±6.3 (1.00x) & 107±20.6 & 108±8.5 (1.00x) & 117±12.0 (0.92x) & 141±12.2 (0.76x) \\ 
 R-chi-square  & 401±1.7 & 402±2.0 (1.00x) & 401±1.8 (1.00x) & 401±3.3 (1.00x) & 3±0.1 & 3±0.1 (0.99x) & 3±0.1 (0.99x) & 3±0.0 (1.00x) \\ 
 C-zxing  & 1224±3.0 & 1223±3.2 (1.00x) & 1223±3.2 (1.00x) & 1222±3.8 (1.00x) & 8±1.2 & 8±1.5 (1.04x) & 7±0.4 (1.11x) & 9±0.9 (0.93x) \\ 
 R-page-rank  & 2702±19.7 & 2701±22.2 (1.00x) & 2703±24.2 (1.00x) & 2694±13.8 (1.00x) & 4±0.1 & 4±0.1 (1.00x) & 4±0.1 (1.01x) & 4±0.1 (1.00x) \\ 
 R-akka-uct  & 6209±33.2 & 6207±48.2 (1.00x) & 6193±45.9 (1.00x) & 6180±37.5 (1.00x) & 3±0.3 & 2±0.3 (1.04x) & 2±0.4 (1.06x) & 3±0.3 (1.00x) \\ 
 \hline Geomean &  & (1.00x) & (1.00x) & (0.99x) &  & (1.02x) & (1.04x) & (1.01x)\\
 \hline\end{tabular}
}
\end{table*}

Table~\ref{tab:eval:sizes} (right) reports the results obtained using {\sf MapReplayBench}, i.e., the replay workloads produced by tracing and replaying the workloads from the DaCapo-Chopin and Renaissance suites.

\paragraph{\textbf{MapReplayBench makes HashMap-relevant performance\\ changes more apparent than application benchmarks.}}
While replay workloads provide potentially many benefits, it is important to establish that they preserve the direction of HashMap-related performance changes observed in the corresponding application benchmarks for the scenarios we study, specifically when it comes to performance behavior.
Given that replay workloads extract  \code{HashMap}-specific events, we expect their performance to be more sensitive to \code{HashMap}-related code changes, enabling better decision making when driving code optimizations.

In general, while performance changes identified using the replay workloads may not be detectable by the application benchmarks (or require prohibitively many runs to detect), we would expect the replay workloads to confirm changes identified by the application benchmarks, and also that the two kinds of benchmarks will not report statistically significant contradictory changes.
Note that because we lack the ground truth, the evaluation is based on statistical hypothesis testing which may produce statistical errors.

Assuming a developer who wants to use the replay workloads to decide on the best value of the DIC parameter, we focus on the 7 benchmarks from the \emph{map-intensive} category that are most relevant to our case study.

\begin{figure}[t]
    \centering
    \caption{Correlation between relative speedups (ratios) in application benchmarks and replay workloads. Different symbols indicate statistical significance of the combinations of speedups. Rectangles indicate the 99\% bootstrap confidence intervals of the corresponding speedup ratios for weakly discordant changes plotted using the $\circ$ symbol.}
    \label{fig:eval:correlation}
    \includegraphics[width=\linewidth,trim=1.9cm 0.90cm 2.4cm 0.95cm, clip]{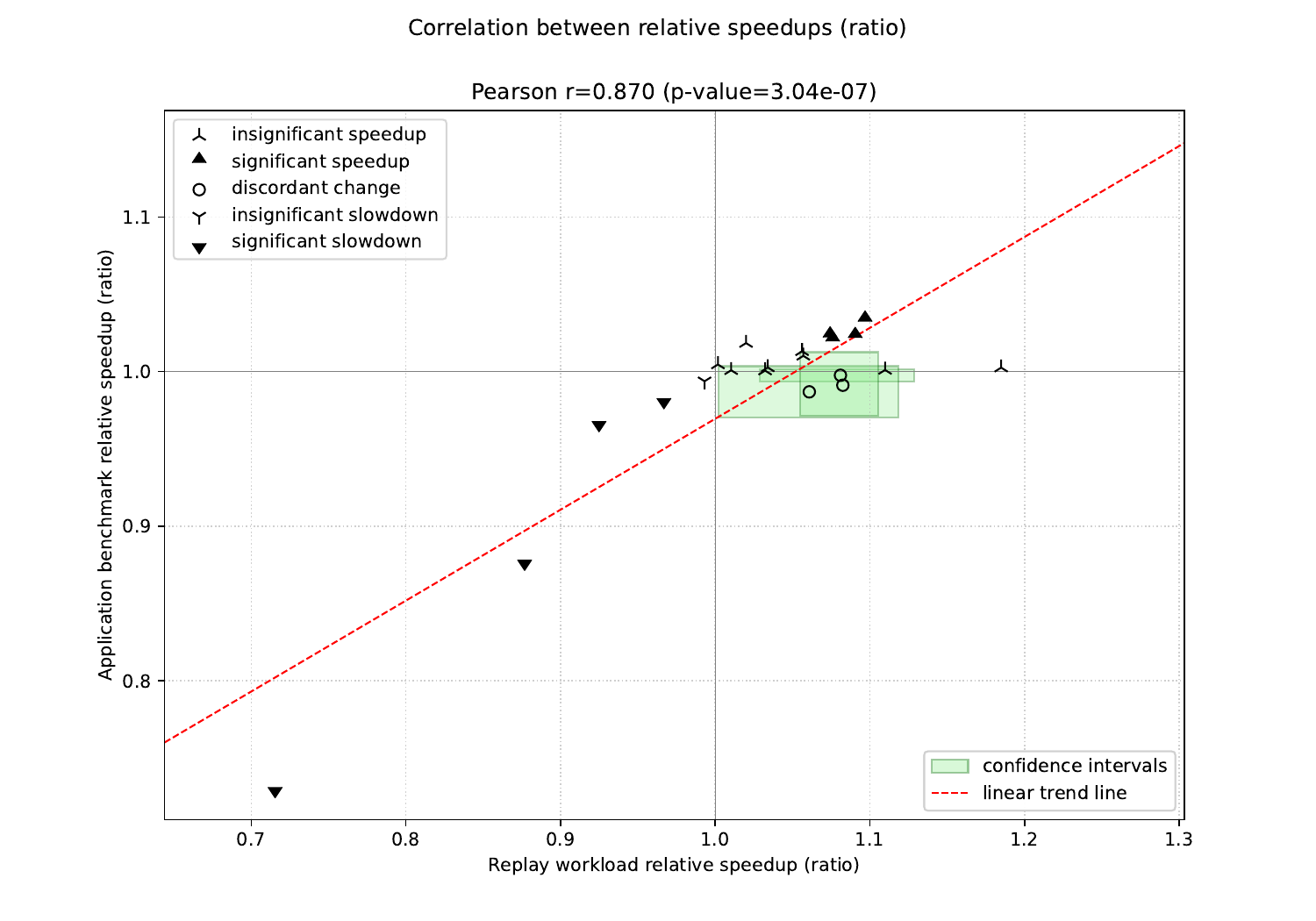}
\end{figure}

We first examine the correlation between changes in the replay workloads and application benchmarks.
Fig.~\ref{fig:eval:correlation} presents a scatter plot of the observed speedup ratios along a linear trend line.
Symbols indicate statistically significant changes, concordant changes (statistically insignificant, same direction), and weakly discordant (at most one statistically significant, opposite directions) changes.\footnote{A performance change (either speedup or slowdown) is considered statistically significant if the 99\% bootstrap confidence interval for the difference of execution time averages excludes zero.
The BCI for each comparison was computed using 50000 bootstrap samples.}
Rectangles illustrate the confidence intervals for the discordant results.

The Pearson correlation coefficient for the ratios is $r = 0.870$, which suggests a strong positive linear relationship.
The extremely low p-value leads us to reject the null hypothesis of no linear relationship with 99\% confidence.
While this aligns with the expectation that the direction and magnitude of performance changes between the two families of benchmarks are related, the correlation does not provide information about specific changes.

To this end, we classify the observed performance changes as shown in Table~\ref{tab:eval:overlap}.
The application benchmarks reported 10 significant performance changes, while the replay workloads reported 18 significant changes.
The overlap group in Table~\ref{tab:eval:overlap} shows that 8 of the significant changes were also reported by the replay workloads, while in the two remaining cases (C-biojava for DIC=32 and C-h2 for DIC=64), the replay workloads indicate a concordant (but not statistically significant) change in the execution time average.
This result is in line with expectations.

The overlap group shows 3 cases of weakly discordant results, i.e., a combination of a statistically significant and insignificant results with opposing directions.
The plot in Fig.~\ref{fig:eval:correlation} provides some indication that these results are close to the boundary and given that the replay-workload results come from only 5 runs (we discuss this later) compared to 30 for the application benchmarks, this may be an artifact of the experimental setup.

While hypothesis testing cannot say anything more about individual statistically insignificant and weakly discordant changes, we can test if observing 18 concordant changes in 21 comparisons could happen by accident.
If the directions of changes in the two benchmark families were independent, observing concordant and discordant changes would be equally likely.

We therefore perform a one-sided binomial test with the null-hypothesis probability $p = 0.5$ for 18 positive outcomes (concordant changes) out of 21 trials.
Given the proportion of positive outcomes of $0.857$, the test reports a p-value of $0.0007$, which allows us to reject the null hypothesis.
In addition, the distance from the proportion of $0.5$ is large (Cohen's $h = 0.796$), we therefore conclude that the directions of changes observed in the replay and application benchmarks are not independent.

Consequently, for the examined use case, we conclude that replay workloads can 
confirm \code{HashMap}-relevant performance changes that are also observable in application benchmarks, but are often less apparent there due to the inclusion of substantial non-\code {HashMap} code.

\paragraph{\textbf{MapReplayBench is more sensitive than application benchmarks.}}
With {\sf MapReplayBench}, we observe clear performance differences even among map-moderate benchmarks---contrasting sharply with the application benchmarks.
While only 7 application benchmarks exhibit measurable differences across DIC configurations, 27 replay workloads show statistically significant changes.

{\sf MapReplayBench} also enables us to identify an ideal DIC value---something neither microbenchmarks nor application benchmarks could reliably reveal.
Microbenchmarks provide insights only on isolated operations, and application benchmarks show minimal sensitivity to DIC changes.
In contrast, {\sf MapReplayBench} indicates that a DIC of 64 yields the highest geometric-mean speedup (1.04×).
We note, however, that our evaluation does not account for the memory footprint associated with larger initial capacities; additional investigation is required to draw a final conclusion.

\paragraph{\textbf{MapReplayBench enables faster feedback than application benchmarks.}}
Given our case study, in choosing the JMH configuration for the replay workloads, we assumed the role of a developer who is familiar with JMH, but does not want to spend much time fine-tuning JMH configurations to specific benchmarks.
In such a situation, choosing to collect data from 5 runs (JMH forks) with 5 warm-up and 5 measurement iterations of 10 seconds each appears to be a sensible choice, accounting for some variability between and within runs.
With this configuration, the experiment execution time for all replay workloads is approximately 8 hours per DIC configuration.\footnote{We note that for some multi-threaded benchmarks (e.g., C-h2, R-scrabble), the original benchmark runs faster than the corresponding replay workload, because the latter is single-threaded.}
Despite the very low number of runs, the replay workloads still reported many statistically significant changes.

\section{Discussion}\label{sec:discussion}

\begin{table}[t]
    \centering    
    \footnotesize
    \setlength{\tabcolsep}{4.75pt}

    \caption{Performance changes with respect to the corresponding baselines with DIC=16, as indicated by the application benchmarks and replay workloads. The $\oplus$ and $\ominus$ symbols indicate statistically significant speedup or slowdown, respectively. The $+$ and $-$ symbols indicate directional shifts in the execution time average, which cannot be considered statistically significant. The overlap combines the indicators from the application and replay workloads. Combinations of two concordant significant changes are preserved, concordant combinations of significant and insignificant changes are reported as insignificant, and discordant changes are reported concatenated using the $|$ symbol.}
    \label{tab:eval:overlap}
    
    \begin{tabular}{l ccc c ccc c ccc}
    \toprule
     & \multicolumn{3}{c}{Replay} &  & \multicolumn{3}{c}{Application} &  & \multicolumn{3}{c}{Overlap} \\
    \cmidrule(lr){2-4} \cmidrule(lr){6-8} \cmidrule(lr){10-12} Benchmark/DIC & 32 & 64 & 128 &  & 32 & 64 & 128 &  & 32 & 64 & 128 \\
    \midrule
    R-par-mnemonics & $\oplus$ & $\oplus$ & $\oplus$ &  & + & $\oplus$ & - &  & + & $\oplus$ & $\oplus$|- \\
    R-mnemonics & $\oplus$ & $\oplus$ & $\oplus$ &  & + & $\oplus$ & - &  & + & $\oplus$ & $\oplus$|- \\
    C-biojava & + & $\oplus$ & $\oplus$ &  & $\oplus$ & $\oplus$ & $\oplus$ &  & + & $\oplus$ & $\oplus$ \\
    R-scrabble & $\ominus$ & $\ominus$ & $\ominus$ &  & $\ominus$ & $\ominus$ & $\ominus$ &  & $\ominus$ & $\ominus$ & $\ominus$ \\
    R-scala-stm-bench7 & $\oplus$ & $\oplus$ & $\oplus$ &  & + & + & + &  & + & + & + \\
    R-fj-kmeans & $\oplus$ & $\oplus$ & $\oplus$ &  & - & + & + &  & $\oplus$|- & + & + \\
    C-h2 & + & - & $\ominus$ &  & + & $\ominus$ & $\ominus$ &  & + & - & $\ominus$ \\
    \bottomrule
    \end{tabular}
\end{table}

This section discusses the broader implications, strengths, and limitations of the {\sf MapReplay} approach.

\subsection{MapReplay as Complementary Benchmarking Approach} \label{sec:complementary}

{\sf MapReplay} is a complementary benchmarking methodology that sits between microbenchmarks and full application benchmarks. Its key advantage is that it extracts \code{HashMap}-relevant activity from real program executions into replay workloads, thereby reducing application-level noise while preserving realistic usage patterns. This enables rapid and reproducible experimentation with configuration choices and implementation variants in a controlled setting. In practice, this is particularly valuable when application benchmarks are costly to run or insufficiently sensitive to \code{HashMap}-level changes, since hash-based map operations often represent only a small fraction of end-to-end execution time in popular Java benchmark suites (as shown in Sec.~\ref{sec:MapReplayBench}).

Replay workloads intentionally omit events unrelated to \code{HashMap} so do not aim to reproduce cross-component effects such as garbage-collection pressure driven by non-map allocations, complex just-in-time compilation interactions spanning the full application, low-level architectural effects influenced by the surrounding code, or multi-threaded contention and synchronization patterns. As a result, {\sf MapReplay} is best suited for studying library-level design choices and performance trade-offs in isolation and for quickly screening hypotheses; application benchmarks remain appropriate to validate end-to-end impact and system-level interactions.

Overall, we position {\sf MapReplay} as a practical middle ground: it retains realistic \code{HashMap} usage derived from applications---often more representative than highly synthetic microbenchmarks---while offering substantially lower experimentation cost and improved actionability when the goal is to understand \code{HashMap}-local behavior.

\subsection{MapReplay Strengths}

{\sf MapReplay} offers several notable strengths that enhance both the practicality and the analytical power of benchmarking hash-based maps. Here, we outline its most important advantages, emphasizing its flexibility, reproducibility, and suitability for detailed performance exploration across a wide range of scenarios.

\paragraph{\textbf{Automated benchmark generation across applications.}}  
A practical advantage of {\sf MapReplay} is that it enables the generation of realistic \code{HashMap} replay workloads from a broad range of Java applications, even when running full end-to-end benchmarks is impractical due to runtime dependencies or environment constraints. 
{\sf MapReplay} captures the relevant \code{HashMap} activity directly during a normal execution of the target application, without any need to reinstantiate any external resource during replay. The resulting replay workload is fully self-contained and can be executed on any JVM, independent of the original application context. This capability makes it possible to produce new, representative benchmarks with minimal effort, greatly expanding the range of real-world scenarios that can be studied under controlled conditions.

\paragraph{\textbf{Flexibility and controlled exploration of design variants.}}
A major strength of {\sf MapReplay} is its ability to provide a controlled and flexible environment for evaluating alternative \code{HashMap}-compatible implementations and design choices. Because directly modifying \code{HashMap} within the Java Class Library is unsafe---due to its deep integration with the JVM---{\sf MapReplay} decouples benchmarking from the JVM’s internal use of the class, enabling recorded traces to be replayed with any hash-based map implementation that adheres to the same API. This includes the standard \code{HashMap}, isolated copies of its implementation (to avoid interference from existing just-in-time compilation profiles), and experimental variants that incorporate complex dependencies or optimizations. At the same time, the replayed benchmark reproduces identical sequences of map operations, allowing researchers to explore design variants such as alternative load factors, hash functions, resizing strategies, or collision-resolution mechanisms under a constant workload. This strict control over both the input and execution context enables fair, apples-to-apples comparisons that are difficult to achieve with conventional application benchmarks, while avoiding JVM instability and ensuring unbiased runtime-compilation behavior. As a result, {\sf MapReplay} serves not only as a safe and extensible platform for isolating and testing new map implementations, but also as a systematic experimental framework for analyzing the sensitivity of \code{HashMap}-like data structures to specific design decisions.

\paragraph{\textbf{Isolation from application-level noise.}}
Our approach reproduces an application's behavior only insofar as it exercises the target library, discarding unrelated computation outside the observed library. By capturing and replaying just the sequence of map operations and their runtime states, {\sf MapReplay} eliminates measurement overhead coming from the rest of the application. This focused replay isolates the library-level effects of an implementation change while preserving the realistic operation mix and state needed to keep measurements representative.

\subsection{MapReplay Limitations and Future Work}\label{sec:limitations}

While {\sf MapReplay} provides a powerful and efficient framework for studying map performance, its design also entails inherent trade-offs and scope limitations. The following discussion examines these constraints, highlighting both their underlying causes and the opportunities they present for future extensions and methodological improvements.

\paragraph{\textbf{Representativeness of captured traces.}}
A central methodological concern lies in the representativeness of the recorded traces. {\sf MapReplay} reproduces the sequence of \code{HashMap} operations captured by the trace (e.g., key properties relevant to control flow) while abstracting away other application effects, but it cannot generalize beyond that specific workload. The performance behavior of \code{HashMap} can vary substantially depending on input characteristics, application phases, and workload intensity. Consequently, if the traced execution reflects only a narrow operational phase or atypical input distribution, the replay workload may not capture the diversity of real-world usage patterns. This limitation is intrinsic to trace-based benchmarking: the fidelity of replay is bounded by the coverage and realism of the collected trace. To mitigate this, one could aggregate traces from multiple representative executions.

\paragraph{\textbf{Impact of just-in-time compilation.}}
A fundamental limitation of our approach arises from the behavior of dynamic (just-in-time) optimizing compilers in multi-tiered JVMs, such as the C2 compiler in HotSpot~VM or Graal in GraalVM. Inlining and optimization decisions can differ significantly between the traced application and the replay workload. This discrepancy is most pronounced for \textit{callbacks}, such as invocations of \code{hashCode()} or \code{equals()} methods, which in real applications may exhibit polymorphism and execute non-trivial user code. In contrast, in our benchmark, callbacks are monomorphic and trivial, consisting of a single basic block. %
While these differences may alter inlining behavior, they generally reduce callback overhead in the replay workload---aligning with our goal to isolate and measure library-level performance rather than application-specific logic.

\paragraph{\textbf{Exclusion of low-level architectural effects.}}
Our approach does not attempt to reproduce low-level architectural phenomena such as cache effects, memory locality, or transient memory pressure. These properties are inherently dependent on the application’s computation patterns and data access behaviors, which are deliberately excluded in our replay. {\sf MapReplay} concentrates execution on the \code{HashMap} library code, abstracting away unrelated application logic. Consequently, while our benchmarks provide high fidelity in the control flow and structural evolution of map operations, they do not reproduce hardware-level performance characteristics. This limitation is acceptable for our focus on evaluating algorithmic and library-level optimizations, but researchers aiming to study microarchitectural effects would require complementary methods.

\paragraph{\textbf{Limitations regarding concurrent maps.}}
Currently, {\sf MapReplay} provides a single-threaded benchmark, even if the traced application is multi-threaded. Our reasoning for correctness relies on the assumption that any shared mutable \code{HashMap} instances in the application are used in a thread-safe manner---typically through synchronization that ensures each map operation is traced atomically within the same critical section as its execution. This assumption guarantees that the traced map-state evolution is consistent with replay. However, this reasoning does not extend to concurrent map implementations that rely on fine-grained synchronization or lock-free mechanisms, such as \code{ConcurrentHashMap}~\cite{lea2005concurrentmap}. 
Extending {\sf MapReplay} to such scenarios is left for future work.

\paragraph{\textbf{Extension to tree-based maps.}}
At present, {\sf MapReplay} focuses on hash-based maps and does not support tree-based data structures such as \code{TreeMap}~\cite{treemap}. Extending our approach to tree maps would require preserving key order information to ensure that mockup keys compare consistently with the original application keys. \code{HashMap} itself occasionally employs treeification of long buckets, converting them into red-black trees; however, such cases are rare in practice. While {\sf MapReplay} correctly operates when treeification occurs, it does not guarantee structural equivalence of treeified buckets between the traced and replayed executions. Achieving faithful replay in these cases would require capturing the comparison relations among keys. Handling ordered or partially ordered maps, as well as treeified \code{HashMap} buckets, remains a topic of ongoing research.

\paragraph{\textbf{Generalization to other collection types.}}
Although \code{HashMap} and its derivative \code{HashSet} are supported, {\sf MapReplay} does not yet address other collection types, such as lists or sets with distinct internal structures. We foresee a main challenge in this direction, collection operations may depend on the state of stored objects, which can be modified by the application. Capturing and replaying all state changes affecting such objects would be prohibitively expensive. A feasible alternative would be to record the sequence of internal execution steps of each operation and synthesize artificial container states that trigger equivalent internal behavior in the benchmark, similarly to our mockup keys. This strategy would also apply to the method \code{HashMap.containsValue()}), which are currently excluded. Generalizing {\sf MapReplay} beyond map-like data structures is conceptually straightforward but practically complex, and thus constitutes an important direction for future work.

\section{Related Work}\label{sec:relwork}

Research on benchmarking, workload reduction, and program analysis provides the methodological foundation for {\sf MapReplay}.  
This section reviews key contributions in six relevant areas: 
(1)~workload distillation and benchmark synthesis,
(2)~record/replay infrastructures,
(3)~program slicing, 
(4)~benchmarking of collection frameworks,
(5)~hash-map algorithms and JVM implementations, and 
(6)~high performance hash tables in research and practice.

\paragraph{\textbf{Workload distillation and benchmark synthesis.}}  
The idea of reducing complex workloads to smaller yet representative benchmarks has been explored in architecture and systems research, e.g., SimPoint~\cite{perelman2013simpoint,10.1145/605397.605403} and MiBench~\cite{guthaus2001mibench}, as well as in compiler and runtime benchmarking~\cite{eeckhout2002quantifying}. 
A complementary line of work synthesizes new, compact benchmarks by matching target workload characteristics: Bell  and John~\cite{bell2005miniaturebenchmarks} propose automatically synthesizing miniature benchmarks that reproduce dynamic metrics of a reference workload; Van Ertvelde and Eeckhout~\cite{vanertvelde2010benchmarksynthesis} synthesize workloads for architecture/compiler exploration by targeting CPU-level characteristics;  Joshi et al.~\cite{Joshi2008Synthetic} argue for the usefulness of synthetic benchmarks and discuss how to construct them for performance studies, while Panda et al.~\cite{panda2017proxy} propose proxy benchmarks to preserve performance characteristics of real-world applications tracking performance metrics  with hardware performance counters. 
{\sf MapReplay} follows a similar principle but focuses on reproducing the control-flow and state transitions of \code{HashMap} operations, achieving realism at the library level without retaining full application logic.
In the database domain, several approaches generate synthesized workloads. Stitcher~\cite{stitcher} synthesizes workloads by combining pieces of standard benchmarks (optionally guided by optimization to match desired performance patterns). 
Similarly, PBench~\cite{pbench} generates synthesized workloads that approximate observed execution statistics. {\sf MapReplay} is designed for a different scope: rather than synthesizing end-to-end workloads to match aggregate performance profiles, it extracts realistic usage patterns for a specific feature of interest (\code{HashMap} operations) from real applications and replays them in a benchmark setting. This  enables focused evaluation of \code{HashMap} optimizations and configuration choices.

\paragraph{\textbf{Record/replay infrastructures.}}  
General-purpose record/replay systems such as PinPlay~\cite{patil2010pinplay} and  DejaVu~\cite{odea2018rr} provide strong determinism for debugging, performance analysis, or fault tolerance. 
These tools typically operate at the process or instruction level and incur substantial overhead due to the volume of recorded events and the need to handle nondeterministic interactions. 
In this line of research, Dolos~\cite{Dolos} focus of record/replay in the context of web application with the ultimate goal of debugging. JaRec~\cite{JaRec} proposes a lightweight Java record/replay infrastructure for multi-threaded applications, again focusing on debugging as final goal.
Similarly rdb~\cite{rdb} is an hardware-assisted record/replay infrastructure for debugging purposes, with strong focus on precise and deterministic replay. 
{\sf MapReplay} differs fundamentally in scope: it records only a strict subset of the application behavior, i.e., \code{HashMap}-related events at the API level, yielding lightweight traces that are fast to replay yet semantically faithful to the traced library behavior.

\paragraph{\textbf{Dynamic program slicing and behavior-preserving extraction.}}  
Program slicing~\cite{weiser1984program} aims to extract the subset of code and data relevant to specific execution behaviors. 
Dynamic slicing~\cite{agrawal1990dynamic,zhang2003dynamic} refines this using runtime dependencies, allowing one to reconstruct only the parts of a program that affect a particular output or side effect. 
Conceptually, {\sf MapReplay} applies a similar reduction: it captures only the execution subset relevant to \code{HashMap} operations, while preserving the internal state transitions of \code{HashMap} objects during replay. Specifically, {\sf MapReplay} applies ideas from the general dynamic slicing techiques to generate workloads that enable quick and reliable performance evaluation of \code{HashMap} variants.

\paragraph{\textbf{Benchmarking collections and empirical studies.}}  

Empirical studies of Java collections show that performance is highly sensitive to usage patterns and workload diversity~\cite{CollectionsBench:ICPE:2017}. Benchmark suites such as DaCapo~Chopin~\cite{Blackburn2006,DaCapo-Chopin:ASPLOS:2025}, Renaissance~\cite{Prokopec2019} and JEDI~\cite{jedi} (a benchmark suite for Java streams generated with S2S~\cite{s2s}) include collection-intensive workloads, but their long runtimes and mixed computation make them expensive when  isolating collection-level effects. Prior work has highlighted pitfalls in microbenchmarking collections, including unrepresentative access distributions and measurement interference~\cite{kalibera2013rigorous}. Repository-mining approaches build domain-specific benchmark suites by automatically extracting and executing unit-test workloads from public repositories~\cite{zheng2016autobench,villazon2019nab}).
 {\sf MapReplay} complements this line of work by enabling fast replay of hash-map activity derived from real application traces. 

JBrainy~\cite{jbrainy}, a Java implementation of Brainy~\cite{brainy}, automatically generates large numbers of synthetic collection microbenchmarks by emitting randomized call sequences. Like {\sf MapReplay}, it supports rapid experimentation with Java \code{HashMaps}; however, it relies on synthetic usage patterns, whereas {\sf MapReplay} extracts realistic patterns from application executions. We expect that {\sf MapReplay}'s application-derived traces lead to insights that better reflect real-world behavior than randomized call sequences.

\paragraph{\textbf{Hash-map algorithms and JVM implementations.}}  
Numerous studies have explored the design and optimization of hash tables, covering open addressing, cuckoo hashing, and cache-efficient variants that improve locality and reduce collision overheads~\cite{pagh2004cuckoo,lim2014efficient}. 
Other work targets concurrent and lock-free hash tables, such as Java’s \code{ConcurrentHashMap}~\cite{lea2005concurrentmap} and follow-up research on scalable non-blocking maps~\cite{click2020concurrenthashmap}. 
These works highlight the performance sensitivity of hash tables and motivate the need for realistic evaluation methodologies such as {\sf MapReplay}.

\paragraph{\textbf{High-performance hash tables in research and practice.}} Hash-table performance is an active research topic with renewed emphasis on modern hardware and memory technologies. Recent work proposes designs that jointly optimize practical criteria (e.g., throughput, memory footprint, and cache behavior)~\cite{bender2023iceberg}, improves vectorized probing strategies across architectures~\cite{boether2023vectorized}, and develops persistent-memory-aware hash tables that optimize writes~\cite{10.5555/3291168.3291202,vogel2022plush}. In industry, major platforms and libraries have introduced substantial hash-table implementation changes driven by robustness and performance, ranging from OpenJDK’s updates to \code{HashMap} collision handling~\cite{openjdkJEP180} to highly optimized production-grade hash tables such as Google's Abseil SwissTables~\cite{abseilSwissTables} and Meta’s F14~\cite{metaF14}. These developments underscore that hash-table optimization is ongoing, motivating methodologies that accurately evaluate performance effects under realistic usage patterns.

\section{Conclusion}\label{sec:conclusion}

This paper presented {\sf MapReplay}, a trace-driven benchmarking methodology for evaluating the performance of \code{HashMap} under application-derived workloads without re-running full applications. {\sf MapReplay} records \code{HashMap}-related API events during execution and replays them through a lightweight replay infrastructure that preserves internal map states for the traced operation sequence. The goal is to preserve performance-relevant \code{HashMap} behavior while enabling reproducible, low-noise performance measurements.

Applying {\sf MapReplay} to DaCapo-Chopin and Renaissance, we constructed {\sf MapReplayBench}, a suite of replay workloads derived from widely used Java benchmarks. We demonstrated our methodology with a case study on determining the fastest default initial capacity (DIC) of \code{HashMap} among four candidates. Our evaluation shows that replay-workload results align with corresponding application benchmarks on the studied map-intensive workloads: across 21 DIC comparisons on 7 workloads, relative speedup ratios exhibited strong positive correlation ($r=0.870$) and most comparisons (18/21) were directionally concordant, showing that replay workloads result in similar performance conclusions as application benchmarks. Moreover, using a practical JMH configuration (5 runs; 5 warmup and 5 measurement iterations of 10s each), replay workloads provided feedback at lower cost than application benchmarks: evaluating all replay workloads for one DIC required about 8 hours, compared to about 72 hours for the corresponding application-benchmark measurements. Replay workloads were more sensitive to \code{HashMap}-local changes: in the same case study they reported 18 statistically significant DIC effects, compared to only 10 for application benchmarks. Across the full benchmark set, only 7 application workloads exhibited measurable performance-relevant DIC differences, while 27 replay workloads showed statistically significant changes. Finally, replay workloads make \code{HashMap}-relevant performance changes more apparent than application benchmarks: they allowed us to identify the DIC yielding the highest speedup (a geometric-mean execution-time improvement of 1.04$\times$), whereas application benchmarks cannot draw such a conclusion.

Considering the above results, we position {\sf MapReplayBench} as a practical middle ground between microbenchmarks and full application benchmarks. It is most useful when the goal is to rapidly and reproducibly evaluate \code{HashMap} configuration choices or implementation variants in a controlled setting, while preserving realistic usage patterns derived from applications and reducing noise from unrelated computation. As replay workloads intentionally abstract away non-\code{HashMap} code and do not aim to capture cross-component effects (e.g., garbage collection driven by non-map allocations, complex JIT interactions spanning the full application, low-level architectural effects influenced by surrounding code, or multi-threaded contention and synchronization), findings from replay workloads should be validated with application benchmarks whenever such end-to-end interactions are expected to matter.

\begin{acks}
This work has been supported by the Hasler Fundation (2025-03-13-355), the Swiss National Science Foundation (IZSEZ0\_229176), and the USI FIR project ``Understanding and Mitigating Performance Variability on Managed Runtimes''.
\end{acks}

\bibliographystyle{ACM-Reference-Format}
\balance
\bibliography{refs}

@inproceedings{CollectionsBench:ICPE:2017,
 author = {Costa, Diego and Andrzejak, Artur and Seboek, Janos and Lo, David},
 title = {Empirical Study of Usage and Performance of Java Collections},
 booktitle = {Proceedings of the 8th ACM/SPEC on International Conference on Performance Engineering},
 series = {ICPE '17},
 year = {2017},
 isbn = {978-1-4503-4404-3},
 location = {L'Aquila, Italy},
 pages = {389--400},
 numpages = {12},
 url = {http://doi.acm.org/10.1145/3030207.3030221},
 doi = {10.1145/3030207.3030221},
 acmid = {3030221},
 publisher = {ACM},
 address = {New York, NY, USA},
}

@ARTICLE{8747433,
  author={Costa, Diego and Bezemer, Cor-Paul and Leitner, Philipp and Andrzejak, Artur},
  journal={IEEE Transactions on Software Engineering}, 
  title={What's Wrong with My Benchmark Results? Studying Bad Practices in JMH Benchmarks}, 
  year={2021},
  volume={47},
  number={7},
  pages={1452-1467},
  doi={10.1109/TSE.2019.2925345}}

@inproceedings{10.1145/2414740.2414747,
author = {Kell, Stephen and Ansaloni, Danilo and Binder, Walter and Marek, Luk\'{a}\v{s}},
title = {The JVM is not observable enough (and what to do about it)},
year = {2012},
isbn = {9781450316330},
publisher = {Association for Computing Machinery},
address = {New York, NY, USA},
url = {https://doi.org/10.1145/2414740.2414747},
doi = {10.1145/2414740.2414747},
booktitle = {Proceedings of the Sixth ACM Workshop on Virtual Machines and Intermediate Languages},
pages = {33–38},
numpages = {6},
location = {Tucson, Arizona, USA},
series = {VMIL '12}
}

@misc{module-patching,
  author       = {{OpenJDK}},
  title        = {{Project Jigsaw: Module System Quick-Start Guide}},
  year         = {2025},
  howpublished = {\url{https://openjdk.org/projects/jigsaw/quick-start}},
}

@inproceedings{10.1145/2162049.2162077,
author = {Marek, Luk\'{a}\v{s} and Villaz\'{o}n, Alex and Zheng, Yudi and Ansaloni, Danilo and Binder, Walter and Qi, Zhengwei},
title = {DiSL: a domain-specific language for bytecode instrumentation},
year = {2012},
isbn = {9781450310925},
publisher = {Association for Computing Machinery},
address = {New York, NY, USA},
url = {https://doi.org/10.1145/2162049.2162077},
doi = {10.1145/2162049.2162077},
booktitle = {Proceedings of the 11th Annual International Conference on Aspect-Oriented Software Development},
pages = {239–250},
numpages = {12},
location = {Potsdam, Germany},
series = {AOSD '12}
}

@manual{JavaHashMapAPI25,
  title        = {Class HashMap},
  author       = {Oracle Corporation},
  organization = {Oracle},
  year         = {2025},
  url          = {https://docs.oracle.com/en/java/javase/25/docs/api//java.base/java/util/HashMap.html},
}

@manual{lea2005concurrentmap,
  author       = {Oracle Corporation},
  title        = {Class ConcurrentHashMap},
  year         = {2025},
  url = {https://docs.oracle.com/en/java/javase/25/docs/api/java.base/java/util/concurrent/ConcurrentHashMap.html},
  }

@inproceedings{10.1145/1960275.1960292,
author = {Moret, Philippe and Binder, Walter and Tanter, \'{E}ric},
title = {Polymorphic bytecode instrumentation},
year = {2011},
isbn = {9781450306058},
publisher = {Association for Computing Machinery},
address = {New York, NY, USA},
url = {https://doi.org/10.1145/1960275.1960292},
doi = {10.1145/1960275.1960292},
booktitle = {Proceedings of the Tenth International Conference on Aspect-Oriented Software Development},
pages = {129–140},
numpages = {12},
location = {Porto de Galinhas, Brazil},
series = {AOSD '11}
}

@article{10.1002/spe.2385,
  author       = {Walter Binder and
                  Philippe Moret and
                  {\'{E}}ric Tanter and
                  Danilo Ansaloni},
  title        = {Polymorphic bytecode instrumentation},
  journal      = {Softw. Pract. Exp.},
  volume       = {46},
  number       = {10},
  pages        = {1351--1380},
  year         = {2016},
  url          = {https://doi.org/10.1002/spe.2385},
  doi          = {10.1002/SPE.2385},
  bibsource    = {dblp computer science bibliography, https://dblp.org}
}

@manual{jni,
  title        = {Java Native Interface Specification},
  author       = {Oracle Corporation},
    url = {https://docs.oracle.com/javase/8/docs/technotes/guides/jni/spec/jniTOC.html},
  year         = {2025},
}

@manual{treemap,
  author       = {Oracle Corporation},
  title        = {Class TreeMap},
  year         = {2025},
  url = {https://docs.oracle.com/en/java/javase/25/docs/api/java.base/java/util/TreeMap.html},
  }

@software{JMH,
  title        = {Java Microbenchmark Harness (JMH)},
  author       = {OpenJDK Community},
  organization = {OpenJDK},
  year         = {2025},
  url          = {https://github.com/openjdk/jmh},
}

@inproceedings{Prokopec2019,
  author    = {Aleksandar Prokopec and Andrea Rosà and David Leopoldseder and Gilles Duboscq and Petr Tůma and Martin Studener and Lubomír Bulej and Yudi Zheng and Alex Villazón and Doug Simon and Thomas Würthinger and Walter Binder},
  title     = {Renaissance: Benchmarking Suite for Parallel Applications on the JVM},
  booktitle = {Proceedings of the 40th ACM SIGPLAN Conference on Programming Language Design and Implementation (PLDI ’19)},
  pages     = {31--47},
  year      = {2019}
}

@inproceedings{Blackburn2006,
author = {Blackburn, Stephen M. and Garner, Robin and Hoffmann, Chris and Khang, Asjad M. and McKinley, Kathryn S. and Bentzur, Rotem and Diwan, Amer and Feinberg, Daniel and Frampton, Daniel and Guyer, Samuel Z. and Hirzel, Martin and Hosking, Antony and Jump, Maria and Lee, Han and Moss, J. Eliot B. and Phansalkar, Aashish and Stefanovi\'{c}, Darko and VanDrunen, Thomas and von Dincklage, Daniel and Wiedermann, Ben},
title = {The DaCapo benchmarks: java benchmarking development and analysis},
year = {2006},
isbn = {1595933484},
publisher = {Association for Computing Machinery},
address = {New York, NY, USA},
url = {https://doi.org/10.1145/1167473.1167488},
doi = {10.1145/1167473.1167488},
booktitle = {Proceedings of the 21st Annual ACM SIGPLAN Conference on Object-Oriented Programming Systems, Languages, and Applications},
pages = {169–190},
numpages = {22},
location = {Portland, Oregon, USA},
series = {OOPSLA '06}
}

@inproceedings{DaCapo-Chopin:ASPLOS:2025,
  author       = {Stephen M Blackburn and
                  Zixian Cai and
                  Rui Chen and
                  Xi Yang and
                  John Zhang and
                  John Zigman
                 },
  title        = {Rethinking {Java} Performance Analysis},
  booktitle    = {Proceedings of the 30th {ACM} International Conference on Architectural
                  Support for Programming Languages and Operating Systems, Volume 1,
                  {ASPLOS} 2025, Rotterdam, Netherlands, 30 March 2025 - 3 April 2025},
  publisher    = {{ACM}},
  year         = {2025},
  url          = {https://doi.org/10.1145/3669940.3707217},
  doi          = {10.1145/3669940.3707217},
}

@article{pagh2004cuckoo,
author = {Pagh, Rasmus and Rodler, Flemming Friche},
title = {Cuckoo hashing},
year = {2004},
issue_date = {May 2004},
publisher = {Academic Press, Inc.},
address = {USA},
volume = {51},
number = {2},
issn = {0196-6774},
url = {https://doi.org/10.1016/j.jalgor.2003.12.002},
doi = {10.1016/j.jalgor.2003.12.002},
journal = {J. Algorithms},
month = may,
pages = {122–144},
numpages = {23}
}

@inproceedings{10.5555/3291168.3291202,
author = {Zuo, Pengfei and Hua, Yu and Wu, Jie},
title = {Write-optimized and high-performance hashing index scheme for persistent memory},
year = {2018},
isbn = {9781931971478},
publisher = {USENIX Association},
address = {USA},
booktitle = {Proceedings of the 13th USENIX Conference on Operating Systems Design and Implementation},
pages = {461–476},
numpages = {16},
location = {Carlsbad, CA, USA},
series = {OSDI'18}
}

@INPROCEEDINGS{lim2014efficient,
  author={Ross, Kenneth A.},
  booktitle={2007 IEEE 23rd International Conference on Data Engineering}, 
  title={Efficient Hash Probes on Modern Processors}, 
  year={2007},
  volume={},
  number={},
  pages={1297-1301},
  doi={10.1109/ICDE.2007.368997}}

@article{click2020concurrenthashmap,
author = {Shalev, Ori and Shavit, Nir},
title = {Split-ordered lists: Lock-free extensible hash tables},
year = {2006},
issue_date = {May 2006},
publisher = {Association for Computing Machinery},
address = {New York, NY, USA},
volume = {53},
number = {3},
issn = {0004-5411},
url = {https://doi.org/10.1145/1147954.1147958},
doi = {10.1145/1147954.1147958},
journal = {J. ACM},
month = may,
pages = {379–405},
numpages = {27}
}

@inproceedings{kalibera2013rigorous,
author = {Kalibera, Tomas and Jones, Richard},
title = {Rigorous benchmarking in reasonable time},
year = {2013},
isbn = {9781450321006},
publisher = {Association for Computing Machinery},
address = {New York, NY, USA},
url = {https://doi.org/10.1145/2491894.2464160},
doi = {10.1145/2491894.2464160},
booktitle = {Proceedings of the 2013 International Symposium on Memory Management},
pages = {63–74},
numpages = {12},
location = {Seattle, Washington, USA},
series = {ISMM '13}
}

@misc{profhot,
  author       = {{Oracle}},
  title        = {Proftool},
  year         = {2025},
  howpublished = {\url{https://github.com/graalvm/mx/blob/master/README-proftool.md}},
}

@INPROCEEDINGS{perelman2013simpoint,
  author={Perelman, E. and Hamerly, G. and Calder, B.},
  booktitle={2003 12th International Conference on Parallel Architectures and Compilation Techniques}, 
  title={Picking statistically valid and early simulation points}, 
  year={2003},
  volume={},
  number={},
  pages={244-255},
  doi={10.1109/PACT.2003.1238020}}

@inproceedings{10.1145/605397.605403,
author = {Sherwood, Timothy and Perelman, Erez and Hamerly, Greg and Calder, Brad},
title = {Automatically characterizing large scale program behavior},
year = {2002},
isbn = {1581135742},
publisher = {Association for Computing Machinery},
address = {New York, NY, USA},
url = {https://doi.org/10.1145/605397.605403},
doi = {10.1145/605397.605403},
booktitle = {Proceedings of the 10th International Conference on Architectural Support for Programming Languages and Operating Systems},
pages = {45–57},
numpages = {13},
location = {San Jose, California},
series = {ASPLOS X}
}

@inproceedings{guthaus2001mibench,
  author={Guthaus, M.R. and Ringenberg, J.S. and Ernst, D. and Austin, T.M. and Mudge, T. and Brown, R.B.},
  booktitle={Proceedings of the Fourth Annual IEEE International Workshop on Workload Characterization. WWC-4 (Cat. No.01EX538)}, 
  title={MiBench: A free, commercially representative embedded benchmark suite}, 
  year={2001},
  volume={},
  number={},
  pages={3-14},
  doi={10.1109/WWC.2001.990739}}

@article{eeckhout2002quantifying,
  title={Quantifying the impact of input data sets on program behavior and its applications},
  author={Eeckhout, Lieven and Vandierendonck, Hans and De Bosschere, Koen},
  journal={Journal of Instruction-Level Parallelism},
  volume={5},
  number={1},
  pages={1--33},
  year={2003}
}

@inproceedings{Joshi2008Synthetic,
  author    = {Ajay M. Joshi and Lieven Eeckhout and Lizy K. John},
  title     = {The Return of Synthetic Benchmarks},
  booktitle = {Proceedings of the 2008 SPEC Benchmark Workshop},
  pages     = {1--11},
  year      = {2008}
}

@article{vogel2022plush,
  author  = {Lukas Vogel and Alexander van Renen and Satoshi Imamura and Jana Giceva and Thomas Neumann and Alfons Kemper},
  title   = {Plush: A Write-Optimized Persistent Log-Structured Hash-Table},
  journal = {Proceedings of the VLDB Endowment},
  volume  = {15},
  number  = {11},
  pages   = {2895--2907},
  year    = {2022},
  doi     = {10.14778/3551793.3551839},
}

@inproceedings{jbrainy,
author = {Couderc, Noric and S\"{o}derberg, Emma and Reichenbach, Christoph},
title = {JBrainy: Micro-benchmarking Java Collections with Interference},
year = {2020},
isbn = {9781450371094},
publisher = {Association for Computing Machinery},
address = {New York, NY, USA},
url = {https://doi.org/10.1145/3375555.3383760},
doi = {10.1145/3375555.3383760},
booktitle = {Companion of the ACM/SPEC International Conference on Performance Engineering},
pages = {42–45},
numpages = {4},
location = {Edmonton AB, Canada},
series = {ICPE '20}
}

@article{brainy,
author = {Jung, Changhee and Rus, Silvius and Railing, Brian P. and Clark, Nathan and Pande, Santosh},
title = {Brainy: effective selection of data structures},
year = {2011},
issue_date = {June 2011},
publisher = {Association for Computing Machinery},
address = {New York, NY, USA},
volume = {46},
number = {6},
issn = {0362-1340},
url = {https://doi.org/10.1145/1993316.1993509},
doi = {10.1145/1993316.1993509},
journal = {SIGPLAN Not.},
month = jun,
pages = {86–97},
numpages = {12}
}

@article{boether2023vectorized,
  author  = {Maximilian B{\"o}ther and Lawrence Benson and Ana Klimovic and Tilmann Rabl},
  title   = {Analyzing Vectorized Hash Tables Across CPU Architectures},
  journal = {Proceedings of the VLDB Endowment},
  volume  = {16},
  number  = {11},
  pages   = {2755--2768},
  year    = {2023},
  doi     = {10.14778/3611479.3611485},
}

@article{bender2023iceberg,
  author  = {Michael A. Bender and Martin Farach-Colton and William Kuszmaul and Giuliano L. Tagliavini},
  title   = {Iceberg Hashing: Optimizing Many Hash-Table Criteria at Once},
  journal = {Journal of the {ACM}},
  year    = {2023},
  doi     = {10.1145/3625817},
}

@inproceedings{stitcher,
  title={Stitcher: Learned Workload Synthesis from Historical Performance Footprints},
  author={Wan, Chengcheng and Zhu, Yiwen and Cahoon, Joyce and Wang, Wenjing and Lin, Katherine and Liu, Sean and Truong, Raymond and Singh, Neetu and Ciortea, Alexandra M and Karanasos, Konstantinos and others},
  booktitle={EDBT},
  pages={417--423},
  year={2023}
}

@inproceedings{10.1145/3578360.3580265,
author = {Basso, Matteo and Ros\`{a}, Andrea and Omini, Luca and Binder, Walter},
title = {Java Vector API: Benchmarking and Performance Analysis},
year = {2023},
isbn = {9798400700880},
publisher = {Association for Computing Machinery},
address = {New York, NY, USA},
url = {https://doi.org/10.1145/3578360.3580265},
doi = {10.1145/3578360.3580265},
booktitle = {Proceedings of the 32nd ACM SIGPLAN International Conference on Compiler Construction},
pages = {1–12},
numpages = {12},
location = {Montr\'{e}al, QC, Canada},
series = {CC 2023}
}

@inproceedings{zheng2016autobench,
  author    = {Yudi Zheng and Andrea Ros{\`a} and Luca Salucci and Haiyang Sun and Omar Javed and Lubom{\'\i}r Bulej and Walter Binder and Yao Li and Zhengwei Qi and Lydia Y. Chen},
  title     = {AutoBench: Finding Workloads That You Need Using Pluggable Hybrid Analyses},
  booktitle = {2016 IEEE 23rd International Conference on Software Analysis, Evolution, and Reengineering (SANER)},
  year      = {2016},
  pages     = {639--643},
  doi       = {10.1109/SANER.2016.70},
  publisher = {IEEE}
}

@inproceedings{villazon2019nab,
  author    = {Alex Villaz{\'o}n and Haiyang Sun and Andrea Ros{\`a} and Eduardo Rosales and Daniele Bonetta and Isabella Defilippis and Sergio Oporto and Walter Binder},
  title     = {Automated Large-Scale Multi-Language Dynamic Program Analysis in the Wild},
  booktitle = {33rd European Conference on Object-Oriented Programming (ECOOP 2019)},
  series    = {Leibniz International Proceedings in Informatics (LIPIcs)},
  year      = {2019},
  doi       = {10.4230/LIPIcs.ECOOP.2019.20}
}

@inproceedings{rosa2018tgp,
  author    = {Andrea Ros{\`a} and Eduardo Rosales and Walter Binder},
  title     = {Analyzing and Optimizing Task Granularity on the {JVM}},
  booktitle = {Proceedings of the 2018 International Symposium on Code Generation and Optimization (CGO)},
  year      = {2018},
  doi       = {10.1145/3168828}
}

@article{rosa2019tgp,
  author    = {Andrea Ros{\`a} and Eduardo Rosales and Walter Binder},
  title     = {Analysis and Optimization of Task Granularity on the Java Virtual Machine},
  journal   = {ACM Transactions on Programming Languages and Systems},
  volume    = {41},
  number    = {3},
  articleno = {19},
  year      = {2019},
  doi       = {10.1145/3338497}
}

@inproceedings{rosa2017rsi,
  author    = {Andrea Ros{\`a} and Eduardo Rosales and Walter Binder},
  title     = {Accurate Reification of Complete Supertype Information for Dynamic Analysis on the {JVM}},
  booktitle = {Proceedings of the 16th ACM SIGPLAN International Conference on Generative Programming: Concepts and Experiences (GPCE)},
  year      = {2017},
  pages     = {216--224},
  doi       = {10.1145/3136040.3136061}
}

@article{rosa2018rsi,
  author  = {Andrea Ros{\`a} and Walter Binder},
  title   = {Optimizing Type-Specific Instrumentation on the {JVM} with Reflective Supertype Information},
  journal = {Journal of Visual Languages and Computing},
  volume  = {49},
  pages   = {29--45},
  year    = {2018},
  doi     = {10.1016/j.jvlc.2018.10.007}
}

@inproceedings{loeff2024vectorizedIntrinsics,
  author    = {J{\'u}nior L{\"o}ff and Filippo Schiavio and Andrea Ros{\`a} and Matteo Basso and Walter Binder},
  title     = {Vectorized Intrinsics Can Be Replaced with Pure Java Code without Impairing Steady-State Performance},
  booktitle = {Proceedings of the 15th ACM/SPEC International Conference on Performance Engineering (ICPE)},
  year      = {2024},
  doi       = {10.1145/3629526.3645051}
}

@article{pbench,
author = {Zhou, Yan and Liu, Chunwei and Urgaonkar, Bhuvan and Wang, Zhengle and Mueller, Magnus and Zhang, Chao and Zhang, Songyue and Pfeil, Pascal and Horn, Dominik and Liu, Zhengchun and Pagano, Davide and Kraska, Tim and Madden, Samuel and Fan, Ju},
title = {PBench: Workload Synthesizer with Real Statistics for Cloud Analytics Benchmarking},
year = {2025},
issue_date = {July 2025},
publisher = {VLDB Endowment},
volume = {18},
number = {11},
issn = {2150-8097},
url = {https://doi.org/10.14778/3749646.3749661},
doi = {10.14778/3749646.3749661},
journal = {Proc. VLDB Endow.},
month = jul,
pages = {3883–3895},
numpages = {13}
}

@misc{openjdkJEP180,
  author       = {{OpenJDK}},
  title        = {{JEP 180}: Handle Frequent {HashMap} Collisions with Balanced Trees},
  howpublished = {\url{https://openjdk.org/jeps/180}},
  year         = {2014},
}

@misc{abseilSwissTables,
  author       = {{Abseil}},
  title        = {Introduction of Swiss Tables in the Abseil {C++} Library},
  howpublished = {\url{https://abseil.io/blog/20180927-swisstables}},
  year         = {2018},
}

@misc{metaF14,
  author       = {{Meta Engineering}},
  title        = {Open-sourcing {F14} for faster, more memory-efficient hash tables},
  howpublished = {\url{https://engineering.fb.com/2019/04/25/developer-tools/f14/}},
  year         = {2019},
}

@inproceedings{bell2005miniaturebenchmarks,
  author    = {Bell Jr., Robert and John, Lizy K.},
  title     = {The Case for Automatic Synthesis of Miniature Benchmarks},
  booktitle = {Workshop on Modeling, Benchmarking and Simulation (MoBS)},
  pages     = {4--8},
  year      = {2005}
}

@inproceedings{vanertvelde2010benchmarksynthesis,
  author    = {Van Ertvelde, Luk and Eeckhout, Lieven},
  title     = {Benchmark Synthesis for Architecture and Compiler Exploration},
  booktitle = {2010 IEEE International Symposium on Workload Characterization (IISWC)},
  pages     = {1--11},
  year      = {2010},
  publisher = {IEEE}
}

@inproceedings{panda2017proxy,
  title={Proxy benchmarks for emerging big-data workloads},
  author={Panda, Reena and John, Lizy Kurian},
  booktitle={2017 26th International Conference on Parallel Architectures and Compilation Techniques (PACT)},
  pages={105--116},
  year={2017},
  organization={IEEE}
}

@ARTICLE{weiser1984program,
  author={Weiser, Mark},
  journal={IEEE Transactions on Software Engineering}, 
  title={Program Slicing}, 
  year={1984},
  volume={SE-10},
  number={4},
  pages={352-357},
  doi={10.1109/TSE.1984.5010248}}

@inproceedings{agrawal1990dynamic,
author = {Agrawal, Hiralal and Horgan, Joseph R.},
title = {Dynamic program slicing},
year = {1990},
isbn = {0897913647},
publisher = {Association for Computing Machinery},
address = {New York, NY, USA},
url = {https://doi.org/10.1145/93542.93576},
doi = {10.1145/93542.93576},
booktitle = {Proceedings of the ACM SIGPLAN 1990 Conference on Programming Language Design and Implementation},
pages = {246–256},
numpages = {11},
location = {White Plains, New York, USA},
series = {PLDI '90}
}

@article{zhang2003dynamic,
  author = {Zhang, Xiangyu and Gupta, Rajiv},
title = {Cost effective dynamic program slicing},
year = {2004},
isbn = {1581138075},
publisher = {Association for Computing Machinery},
address = {New York, NY, USA},
url = {https://doi.org/10.1145/996841.996855},
doi = {10.1145/996841.996855},
booktitle = {Proceedings of the ACM SIGPLAN 2004 Conference on Programming Language Design and Implementation},
pages = {94–106},
numpages = {13},
location = {Washington DC, USA},
series = {PLDI '04}
}

@inproceedings{patil2010pinplay,
author = {Patil, Harish and Pereira, Cristiano and Stallcup, Mack and Lueck, Gregory and Cownie, James},
title = {PinPlay: a framework for deterministic replay and reproducible analysis of parallel programs},
year = {2010},
isbn = {9781605586359},
publisher = {Association for Computing Machinery},
address = {New York, NY, USA},
url = {https://doi.org/10.1145/1772954.1772958},
doi = {10.1145/1772954.1772958},
booktitle = {Proceedings of the 8th Annual IEEE/ACM International Symposium on Code Generation and Optimization},
pages = {2–11},
numpages = {10},
location = {Toronto, Ontario, Canada},
series = {CGO '10}
}

@inproceedings{odea2018rr,
  title={Deterministic replay of Java multithreaded applications},
  author={Choi, Jong-Deok and Srinivasan, Harini},
  booktitle={Proceedings of the SIGMETRICS symposium on Parallel and distributed tools},
  pages={48--59},
  year={1998}
}

@article{boncz2020fsst,
  title={{FSST: fast random access string compression}},
  author={Boncz, Peter and Neumann, Thomas and Leis, Viktor},
  journal={Proceedings of the VLDB Endowment},
  volume={13},
  number={12},
  pages={2649--2661},
  year={2020},
  publisher={VLDB Endowment}
}

@inproceedings{jiang2021good,
  title={{Good to the last bit: Data-driven encoding with codecdb}},
  author={Jiang, Hao and Liu, Chunwei and Paparrizos, John and Chien, Andrew A and Ma, Jihong and Elmore, Aaron J},
  booktitle={Proceedings of the 2021 International Conference on Management of Data},
  pages={843--856},
  year={2021}
}

@article{JaRec,
  title={JaRec: a portable record/replay environment for multi-threaded Java applications},
  author={Georges, Andy and Christiaens, Mark and Ronsse, Michiel and De Bosschere, Koenraad},
  journal={Software: practice and experience},
  volume={34},
  number={6},
  pages={523--547},
  year={2004},
  publisher={Wiley Online Library}
}

@inproceedings{Dolos,
  title={Interactive record/replay for web application debugging},
  author={Burg, Brian and Bailey, Richard and Ko, Amy J and Ernst, Michael D},
  booktitle={Proceedings of the 26th annual ACM symposium on User interface software and technology},
  pages={473--484},
  year={2013}
}

@article{rdb,
  title={Replay debugging: Leveraging record and replay for program debugging},
  author={Honarmand, Nima and Torrellas, Josep},
  journal={ACM SIGARCH Computer Architecture News},
  volume={42},
  number={3},
  pages={445--456},
  year={2014},
  publisher={ACM New York, NY, USA}
}

@inproceedings{jedi,
	Author = {Filippo Schiavio and Walter Binder},
  booktitle={Proceedings of The 48th International Conference on Software Engineering},
	Title = {{JEDI}: {Java} Evaluation of Declarative and Imperative Queries -- Benchmarking the {Java Stream API}},
	Year = {2026},
    doi = {10.1145/3744916.3773165},
}

@inproceedings{misleading-micro,
	Author = {Filippo Schiavio and Lubomír Bulej and Walter Binder},
  booktitle={Proceedings of The 41st ACM/SIGAPP Symposium on Applied Computing},
	Title = {{Misleading Microbenchmarks on the Java Virtual Machines}},
	Year = {2026},
    doi = {10.1145/3748522.3779882},
}

@inproceedings{s2s,
  author    = {Filippo Schiavio and
               Andrea Ros{\`{a}} and
               Walter Binder},
  -editor    = {Bernhard Scholz and
               Yukiyoshi Kameyama},
  title     = {{{SQL} to Stream with {S2S:} An Automatic Benchmark Generator for the
               Java Stream {API}}},
  booktitle = {Proc. 21st ACM/SIGPLAN Intl. Conf. on Generative Programming: Concepts and Experiences},
  series    = {GPCE},
  pages     = {179--186},
  publisher = {ACM},
  year      = {2022},
  url       = {https://doi.org/10.1145/3564719.3568699},
  doi       = {10.1145/3564719.3568699},
  timestamp = {Mon, 05 Dec 2022 09:56:37 +0100},
  biburl    = {https://dblp.org/rec/conf/gpce/SchiavioRB22.bib},
  bibsource = {dblp computer science bibliography, https://dblp.org}
}

\appendix

\end{document}